\documentclass{IEEEtran}
\IEEEoverridecommandlockouts

\usepackage{cite}
\usepackage{amsmath,amssymb,amsfonts}
\usepackage{algorithmic}
\usepackage{graphicx}
\usepackage{textcomp}
\usepackage{subcaption}
\usepackage{xcolor}
\usepackage{placeins}
\usepackage[caption=false]{subfig}
\def\BibTeX{{\rm B\kern-.05em{\sc i\kern-.025em b}\kern-.08em
    T\kern-.1667em\lower.7ex\hbox{E}\kern-.125emX}}

\label{English Chars}

\newcommand{\bv}{{\bf v}}

\newcommand{\ju}{\mathrm{j}}

\newcommand{\tx}{\mathrm{tx}}
\newcommand{\rx}{\mathrm{rx}}

\begin{document}

\title{Multiple-Input Multiple-Output Forward Scatter Radar \\
\thanks{}
}

\author{\IEEEauthorblockN{Abdollah Ajorloo, Stefano Amato Ciciretti, and Fabiola Colone}\\
\IEEEauthorblockA{\textit{DIET Department}, \textit{Sapienza University of Rome}, Rome, Italy \\
\{abdollah.ajorloo,stefanoamato.ciciretti,fabiola.colone\}@uniroma1.it}
}

\maketitle

\begin{abstract}
Forward Scatter Radar (FSR) offers attractive capabilities for the detection of low-observable targets by exploiting the enhanced radar cross section that arises when a target crosses the transmitter–receiver baseline. While conventional FSR systems are typically implemented using single-input single-output (SISO) architectures and provide limited localization capabilities, recent developments in array-based processing have enabled non-coherent angular estimation through single-input multiple-output (SIMO) configurations. In this paper, we introduce a comprehensive Multiple-Input Multiple-Output Forward Scatter Radar (MIMO-FSR) framework that enables joint estimation of the target Direction of Departure (DoD) and Direction of Arrival (DoA) using non-coherent, amplitude-only measurements. The proposed approach extends the spatial-domain processing paradigm of FSR by exploiting antenna arrays at both the transmitter (TX) and receiver (RX) sides, allowing two-dimensional angular localization without requiring phase synchronization among TX/RX array elements. We first establish the fundamental principle of operation under ideal signal separability and subsequently develop practical implementations based on Frequency-Division Multiplexing (FDM) and Time-Division Multiplexing (TDM). For each strategy, dedicated signal models and processing schemes are derived to compensate for the distortions introduced by frequency offsets and target motion, respectively. Analytical expressions describing angular resolution, ambiguity conditions, and output map distortions are also provided. Extensive numerical simulations are conducted for both short- and long-baseline scenarios, highlighting the operational limits and trade-offs of the two multiplexing schemes. Finally, proof-of-concept experimental campaigns using software-defined radio platforms validate the proposed framework in real outdoor environments. The results demonstrate successful joint DoD–DoA estimation and confirm the feasibility of MIMO-FSR as a flexible and scalable architecture for future active, passive, and integrated sensing and communications applications.
\end{abstract}


\section{Introduction}

Forward Scatter Radar (FSR) is a specialized class of bistatic radar operating with a bistatic angle approaching $180^\circ$. Target detection in FSR relies on the forward scattering effect, which occurs when an object moves around and intercepts the baseline or line-of-sight between the transmitter (TX) and the receiver (RX). This creates a shadow contour that interrupts the direct path signal, generating a distinct and detectable modulation in the received signal amplitude. Despite its lack of range and velocity resolution in the classical radar sense, FSR boasts compelling operational advantages, including a significant and stable enhancement in the forward scattering radar cross-section (RCS) and an exceptionally simple hardware and signal processing architecture, which enables the detection of stealth or low-RCS targets \cite{cherniakov2007basic, gashinova2013phenomenology, Gashinova2017, Glaser1985, Willis2005, contu2017passive, ustalli2017detection}. Furthermore, FSR is uniquely suited for integration into passive radar (PR) networks and Integrated Sensing and Communications (ISAC) frameworks, allowing the opportunistic or cooperative exploitation of existing commercial transmissions---such as GSM, DVB-T, or GNSS---to expand surveillance coverage with a minimal spectrum footprint \cite{contu2017passive, colone2020dvb, krysik2013gsm, ilioudis2019gnss}.

Traditionally, FSR systems have been deployed in Single-Input Single-Output (SISO) configurations. These architectures rely on time-domain processing to capture the target-induced amplitude fluctuations across time as a target traverses the baseline \cite{cherniakov2007basic, ustalli2017detection, contu2017passive, colone2020dvb}. While multi-static FSR configurations have been explored to reconstruct target trajectories and estimate kinematic parameters \cite{pastina2016target, ustalli2019target}, detection in these systems is still conventionally executed by applying independent time-domain processing at individual receiver nodes before fusing the extracted target Doppler signatures. 

Alternatively, some studies have introduced Single-Input Multiple-Output (SIMO) FSR systems that leverage spatial-domain array processing \cite{Ajorloo2025Multil, myakinkov2003space, Qin2024, qin2025music,11046028}. By analyzing spatial variations of the received amplitude across an antenna array, this non-coherent, amplitude-only paradigm allows for Direction of Arrival (DoA) estimation for the observed target. Crucially, because it operates without requiring absolute phase synchronization across elements, this approach offers robustness to phase noise; moreover, it exhibits flexibility regarding the underlying transmit waveform \cite{Ajorloo2025Multil, Qin2024, ajorloo2024preliminary}.

Building on these spatial-domain insights and extending our preliminary conference study \cite{ajorloo2025forward}, this paper presents a comprehensive framework for Multiple-Input Multiple-Output (MIMO) Forward Scatter Radar. Specifically, we show that, dual to the SIMO configuration, the received signal exhibits a spatial amplitude modulation with respect to the transmit antenna index. In particular, for a fixed receive antenna, the amplitudes associated with the different transmit array elements form a spatial pattern analogous to that observed across the receive array in SIMO FSR. By exploiting multiple antennas at both ends of the baseline, the proposed MIMO-FSR framework enables a powerful new capability: the joint estimation of the target’s Direction of Departure (DoD) and DoA using solely the instantaneous amplitudes of the received signals. This two-dimensional angular localization provides a major step forward over the 1D DoA estimation restricted to SIMO configurations. Additionally, the architecture enables Multiple-Input Single-Output (MISO) sub-configuration, allowing cost-effective, single-antenna receiver deployed arbitrarily, to perform transmit-side angular estimation. It can also accommodate distributed MISO configurations, using multiple single-antenna receivers deployed at diverse locations, to further increase the situational awareness, combining also the conventional time-domain processing.   

The key challenge in enabling transmit diversity in an amplitude-only FSR system is separating the contributions from different transmitters while preserving the target-induced amplitude-modulation profile. In conventional MIMO radar, this is typically achieved through orthogonal waveform multiplexing. However, this approach is not well suited to MIMO-FSR, as it generally relies on coherent, phase-sensitive processing, increasing receiver complexity and imposing more stringent hardware requirements. This would undermine one of the main advantages of amplitude-only FSR architectures: their low hardware complexity. To introduce the operating principle of the proposed 2D processing scheme, we first consider an idealized scenario in which the transmitted signals are perfectly separable despite occupying the same time–frequency slot. We then address the separability challenge through two practical signal-division strategies—Frequency-Division Multiplexing (FDM) and Time-Division Multiplexing (TDM)—and develop dedicated processing schemes for each.

Under the FDM approach, discrete carrier frequency offsets are assigned to individual transmit channels, enabling band-selective pre-filtering at each receiver to isolate the signals prior to amplitude extraction. 
However, this frequency offset causes an unwanted distortion in the amplitude modulation across the transmit antenna index, leading to an angular mispositioning. Specifically, the distortion depends on the unknown bistatic range, thereby introducing an undesirable range--angle coupling. To compensate for this effect and recover the correct target location, the bistatic range is replaced by its closed-form expression derived from the baseline geometry, which expresses it as a function of the target angles.
This maps the DoD and DoA parameters into a warped coordinate grid, which inherently introduces non-uniform and asymmetric modifications to the DoA-DoD map. 

Conversely, under the TDM strategy, transmitters operate sequentially in distinct time slots to ensure single-source reception at any given instant. While TDM avoids frequency-dependent range coupling, the target's physical displacement across consecutive slots introduces a motion-induced distortion in the received signal amplitude proportional to the bistatic range rate (velocity). This induces a distinct shift along the DoD direction. We counter this distortion by implementing either a parallel SISO Doppler estimation path or a fully integrated 3D processing chain (DoD-DoA-Doppler) that simultaneously solves for the spatial frequencies and resolves target velocity sign ambiguities. 

To evaluate the operational boundaries and robustness of the proposed MIMO-FSR architectures, we conduct high-fidelity numerical simulations across both short-baseline scenarios (e.g., for people and ground vehicles monitoring) and long-baseline configurations (e.g., for UAV or aircraft surveillance). These simulations quantify the impact of uncompensated angular biases, tracking update rates, and asymmetric resolution variations induced by the FDM and TDM corrections. 

More importantly, the theoretical frameworks are validated through real-world outdoor experimental campaigns utilizing software-defined radio (SDR) platforms. Some preliminary experimental results is reported in our conference paper \cite{Ajorloo2026Eurad}. 
For the FDM campaign, an array configuration consisting of $M=4$ transmit and $N=8$ receive elements is deployed over a 37-meter baseline to track a walking human target using 16-QAM OFDM signals, demonstrating successful 2D angular localization. For the TDM campaign, the processing chain is validated using an emulated stationary-target scenario, in which a single transmit antenna is mechanically displaced to synthesize an antenna array. The validation is performed with both sinusoidal and OFDM waveforms. 

The key contributions of this work are summarized as follows:
\begin{itemize}
    \item \textbf{Amplitude-only MIMO-FSR framework:}  A comprehensive non-coherent MIMO-FSR framework is introduced for the joint estimation of the target’s DoD and DoA from instantaneous amplitude-only measurements. The framework extends spatial-domain SIMO-FSR processing to arrays deployed at both ends of the baseline and also supports MISO configurations.
    \item \textbf{Practical multiplexing and distortion compensation:} FDM- and TDM-based implementations are developed to separate the transmit-channel contributions without requiring coherent, phase-sensitive processing. For both the FDM and TDM approaches, we investigate the impact of the signal-separation mechanisms on the achievable result and we present the specialized processing schemes with applied corrections.
    \item \textbf{Theoretical and numerical performance assessment:} The angular resolution, ambiguity conditions, and residual distortions associated with both multiplexing strategies are analyzed. Extensive numerical simulations covering short- and long-baseline scenarios are then used to evaluate the operational limits, robustness, and trade-offs of the proposed FDM and TDM architectures. Also the MISO scalability is discussed along with its practical advantages
    \item \textbf{Experimental validation:} Outdoor proof-of-concept campaigns using SDR platforms validate the proposed processing framework with sinusoidal and OFDM waveforms, demonstrating the feasibility of joint DoD–DoA estimation under realistic operating conditions..
\end{itemize}

The remainder of this paper is organized as follows: Section~\ref{sec:SystemModel} establishes the system geometry. Section~\ref{SecMimoPrinciple} presents the fundamental principle of operation of MIMO-FSR under idealized signal separability. Sections~\ref{sec:FDM} and \ref{sec:TDM} develop the practical separation strategies using FDM and TDM, respectively. Section~\ref{sec:Simulations} presents comprehensive numerical simulation results for short- and long-baseline scenarios. Section~\ref{sec:experimental_results} details the real-world verification via outdoor experimental data collected for both architectures. Finally, Section~\ref{sec:Conclusion} concludes the paper.

\section{System Geometry and Signal Model}\label{sec:SystemModel}
\begin{figure}[!t]
	\centering
\includegraphics[width = \columnwidth]{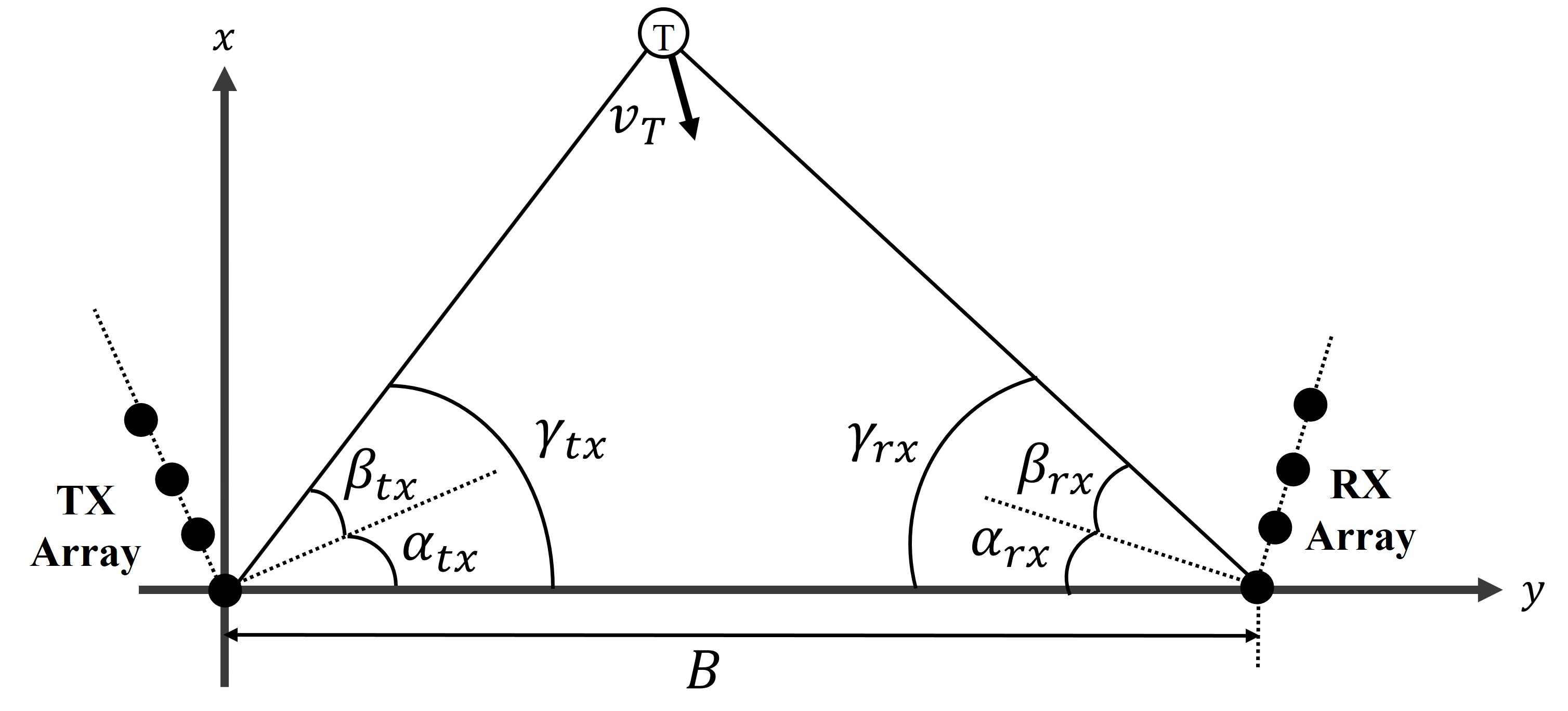}
	\caption{MIMO FSR Geometry}
    	\label{FigGeo}
\end{figure}
System geometry is illustrated in Fig. \ref{FigGeo}.  
We consider a MIMO configuration featuring a transmit (TX) array with \( M \) antennas and a receive (RX) array with \( N \) antennas. The TX and RX arrays are assumed to be located in the far-field of each other, separated by a baseline \( B \). The first TX antenna is positioned at the origin \((0, 0)\) of the Cartesian coordinate system. The remaining TX elements are placed at distances \( d_{tx,m} \), for \( m = 0, \ldots, M-1 \), measured relative to the first TX element (with \( d_{tx,0} = 0 \)). Similarly, the RX array is anchored at \((B, 0)\), with the distance of the \( n \)th RX element given by \( d_{rx,n} \), relative to the first RX element (\( d_{rx,0} = 0 \)). The TX and RX arrays are squinted by angles \( \alpha_{tx} \) and \( \alpha_{rx} \), respectively, with respect to the baseline as shown in the figure. We define $B_{mn}$ as the baseline distance between the $m$th TX and $n$th RX antennas. It follows that
\begin{equation}
	B_{mn} = B + d_{tx,m}\sin \alpha_{tx} + d_{rx,n}\sin \alpha_{rx} ~~_{\substack{m=0,\ldots,M-1\\ n=0,\ldots,N-1}}. 
\end{equation}
A prospective target is assumed to traverse the baseline between the TX and RX arrays. As the target moves along its path, the line connecting the first TX antenna to the target forms an angle \( \gamma_{tx}(t) \) with the baseline, while the line connecting the first RX antenna to the target forms an angle \( \gamma_{rx}(t) \), as illustrated in Fig. \ref{FigGeo}. The distances from the target to the \( m \)th TX and the \( n \)th RX antennas are denoted by \( R_{tx,m}(t) \) and \( R_{rx,n}(t) \), respectively. Assuming that the target remains in the far-field region of both the TX and RX arrays, the following equations hold:
\begin{equation}
	\begin{aligned}
		R_{tx,m}(t) &= R_{tx,0}(t) -  d_{tx,m}\sin\beta_{tx}(t),~~ {\scriptstyle m=0,\ldots,M-1},\\
		R_{rx,n}(t) &= R_{rx,0}(t) -  d_{rx,n}\sin\beta_{rx}(t),~~ {\scriptstyle n=0,\ldots,N-1},
	\end{aligned}
\end{equation}  
where $\beta_{tx}(t) = \gamma_{tx}(t)-\alpha_{tx}$ and $\beta_{rx}(t) =\gamma_{rx}(t)-\alpha_{rx}$ represent DoD and DoA of target, respectively.


The $m$th TX antenna transmits a narrow-band unitary power level signal denoted by $s_m(t)$ in baseband. Consequently, the baseband signal received at the $n$th RX element due to illumination by the $m$th TX element can be expressed as
\cite{ustalli2017detection, colone2020dvb, Ajorloo2025Multil}:

\begin{align}
	x_{mn}(t) &=  A_{D}^{} s_{m}(t)  e^{j\phi_{D,mn}^{}}\qquad\qquad{\substack{m=0,\ldots,M-1\\ n=0,\ldots,N-1}}\\ 
	&+ A_{T}^{}(t) s_{m}(t-\tau_{\rm bi}(t))  e^{j\phi_{T,mn}^{}(t)} + \zeta_{mn}(t),\nonumber
\end{align}
where $A_{D}^{}$ is the direct path amplitude accounting for the path loss between the TX and RX arrays and we assume the $ M $ transmitters to be of equal power level, $\phi_{D,mn}^{}$ is the phase difference in the direct path from TX antenna $m$ to RX antenna $n$ ($\phi_{D,00}^{} = 0$). Further, $A_{T}^{}(t)$, stands for the amplitude in the indirect path encapsulating the TX-target-RX path loss along with the target forward-scattering pattern, as modeled in \cite{ustalli2017detection}, $\tau_{\rm bi}(t)$ is the bistatic delay, and $\phi_{T,mn}^{}(t)$ is the phase difference in the indirect path from TX antenna $m$ to target to RX antenna $n$. Moreover, $ \zeta_{mn}(t)$ denotes the additive white Gaussian noise, assumed to
be statistically independent for distinct index pairs (m, n).. The following equations are held: 
\begin{align}
		\phi_{D,mn}^{} &= -\frac{2\pi}{\lambda}[B_{mn} - B]\label{eqDirPhase}\\ 
		&= -\frac{2\pi}{\lambda}\left [ d_{tx,m}\sin \alpha_{tx} + d_{rx,n}\sin \alpha_{rx}  \right] ,\nonumber\label{eqTarPhase}\\
		\phi_{T,mn}^{}(t) &= -\frac{2\pi}{\lambda}\left[ R_{tx,m}(t)+R_{rx,n}(t)- B \right]\\
			&= \phi_{T,00}^{}(t) + \frac{2\pi}{\lambda}\left [d_{tx,m}\sin \beta_{tx} + d_{rx,n}\sin \beta_{rx}\right ],\nonumber\\
			\tau_{\rm bi}(t) &=\frac{1}{c}R_{\rm bi}(t) = \frac{1}{c}[R_{tx,0}(t)+R_{rx,0}(t)-B].			
\end{align}

Therefore, the composite signal received at the $n$th Rx element is expressed as the sum of the contributions resulting from the $M$ transmitters:
\begin{equation}\label{eq_xn}
	x_n(t) = \sum_{m=0}^{M-1} x_{mn}(t).
\end{equation}
\subsection*{Single Transmitter Case:}
When a single TX is present (i.e., $M=1$), as in SISO and SIMO configurations, the principle of FSR is easily exploited by extracting the amplitude of the signal received at each RX:
\begin{equation}
    \label{eqSqMod}
	y_{0n}(t)=|x_{0n}(t)|^2,\quad   
    n=0,..,N-1
\end{equation}
Based on this input, conventional SISO FSR processing exploits the temporal amplitude modulation generated at a single RX channel as the target crosses the baseline, thereby detecting the target in the Doppler domain and providing information on its bistatic velocity \cite{cherniakov2007basic, ustalli2017detection, contu2017passive, colone2020dvb}. In a SIMO configuration, the target also induces an amplitude modulation across the receive array whose spatial frequency depends on its DoA. Accordingly, spatial-domain processing of $y_{0n}(t)$ across the antenna index $n$ enables target detection and angular localization \cite{Ajorloo2025Multil, myakinkov2003space, Qin2024}. 

To illustrate the inherent principle, we neglect noise and assume the typical condition $A_T(t)\ll A_D$, so that eq. \eqref{eqSqMod} can be approximated as
\begin{align}\label{eqAmpTgt}
y_{0n}(t)
&\simeq
\left|A_Ds_0(t)\right|^2
\nonumber\\
&
+2\Re\left\{
A_Ds_0(t)A_T^{*}(t)s_0^{*}(t-\tau_{\rm bi}(t))
\mathrm{e}^{j[\phi_{D,0n}-\phi_{T,0n}(t)]}
\right\}.
\end{align}
The first term represents the direct-path contribution and is typically removed by a DC-removal stage. This can be applied in the temporal domain, which is effective when a constant amplitude waveform is exploited \cite{colone2020dvb}, or in the spatial domain, which preserves stationary targets \cite{Ajorloo2025Multil}. 
Using \eqref{eqDirPhase} and \eqref{eqTarPhase}, the residual target-dependent component can be rewritten as
\begin{subequations}
\addtocounter{equation}{-1}
 \begin{equation}\label{eqSIMOAmp}
     w_{0n}(t) = \vert A_0(t)\vert \cos(\phi_{T,00}(t) + \tilde{\phi}_n(t) + \angle A_{0}(t))    
 \end{equation}
\begin{align}
    \phi_{T,00}(t) &= -\frac{2\pi}{\lambda}\left[ R_{\tx,0}(t)+R_{\rx,0}(t)- B \right] \label{eq:suba} \\
    \tilde{\phi}_n(t) &= \frac{2\pi}{\lambda} d_{rx,n} \left[ \sin\beta_\rx(t)+\sin\alpha_\rx\right] 
\end{align} 
 \end{subequations} 
where
$A_0(t)= 2A_D^{*}A_T(t) s_0^{*}(t)s_0(t-\tau_{\rm bi}(t))$. 
Equation \eqref{eqSIMOAmp} explicitly separates the amplitude modulation mechanisms exploited by SISO and SIMO configurations, respectively. The common phase term $\phi_{T,00}(t)$, which varies with the bistatic range, governs the temporal amplitude modulation exploited by conventional SISO processing. In this case, an FFT across temporal samples allows the detection of moving targets and their bistatic Doppler estimation. In contrast, $\tilde{\phi}_n(t)$ introduces a phase progression across the receive-array index and therefore determines the spatial amplitude modulation exploited for DoA estimation in SIMO FSR. Target detection and its DoA localization can be obtained via an FFT performed across the spatial samples at individual snapshots. 

\vspace{0.3cm}
In the following, we show that an analogous target-induced amplitude modulation also arises across the transmit-array index $m$. This observation provides the fundamental basis for extending the amplitude-only FSR processing framework to MISO and MIMO configurations.
However, in a MIMO configuration, if all transmitters radiate simultaneously over the same time--frequency resources, the amplitude modulation associated with each individual transmitter cannot be directly extracted from the composite received signal in \eqref{eq_xn}.
In conventional MIMO radar, this issue is typically addressed by employing orthogonal waveforms together with coherent, phase-sensitive processing. Such processing, however, must be performed prior to amplitude extraction, thereby increasing the receiver complexity and compromising the inherent simplicity of amplitude-only FSR. Therefore, we first introduce the proposed MIMO-FSR framework under the ideal assumption of perfect signal separability, namely that the contribution of each transmitter in \eqref{eq_xn} is individually available irrespective of the transmitted waveforms. Subsequently, in Sections~\ref{sec:FDM} and~\ref{sec:TDM}, we address the practical signal-separation problem by developing two implementations based on Frequency-Division Multiplexing (FDM) and Time-Division Multiplexing (TDM), respectively.

\section{MIMO Principle of Operation under Ideal Separability}
\label{SecMimoPrinciple}
\begin{figure}[!t]
	\centering
\includegraphics[width = \columnwidth]{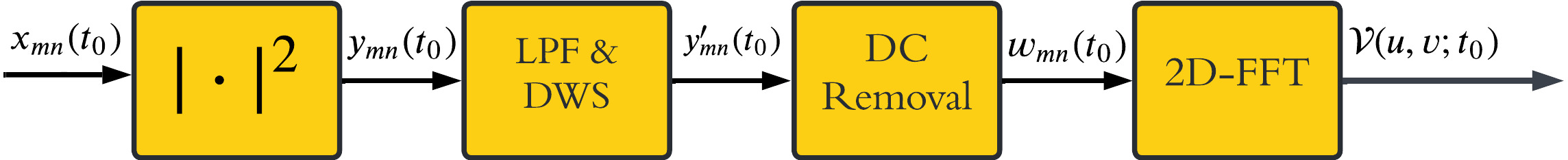}
	\caption{processing chain for an individual TX-RX channel in a MIMO apprach}
    	\label{FigProcChain1Channel}
\end{figure}

In order to introduce the principle of operation of a MIMO-FSR, we first consider an ideal case where it is assumed that the contributions due to different TX array elements can be separated at the RX side even when all signals are transmitted simultaneously and share the same frequency band. 
In other words, we assume that the amplitude of each individual signal in \eqref{eq_xn} is available at the receiver:
\begin{equation}
	y_{mn}(t)=|x_{mn}(t)|^2 ,  \  
    n=0,..,N-1
\end{equation}
Each signal component, is then subject to a simple processing chain as depicted in Fig. \ref{FigProcChain1Channel}, similar to that used in conventional FSR systems. Each amplitude signal endures low-pass filtering to confine the possible amplitude fluctuations due to the transmitting signals ($s_m(t)$) and retain a small portion of the frequency band corresponding to the maximum expected bistatic Doppler shift for targets of interest. This also paves the way for a significant reduction of the sampling rate as the processing chain is typically implemented in digital domain. Specifically, we have
\begin{subequations}\label{eqSigIdealTot}    
\begin{flalign}
\tilde{y}_{mn}(t) &= \mathrm{LPF}\{y_{mn}(t)\},&&
\end{flalign}
\vspace{-0.7cm}
\begin{flalign}\label{eqSigIdeal}
	y_{mn}&(t) \simeq |A_{D}s_{m}(t)|^2 + |A_{T}(t)s_{m}(t-\tau_{\rm bi}(t))|^2\\ 
	+& 2 \Re\left\{A_{D}^{}A_{T}^{*}(t) s_{m}(t)s_{m}^{*}(t-\tau_{\rm bi}(t)) \mathrm{e}^{j\left[\phi_{D,mn}^{}-\phi_{T,mn}^{}(t)\right]} \right\}.\nonumber
\end{flalign}
\end{subequations}
where $\mathrm{LPF}\{\star\}$ denotes the low-pass filtering and, as in the previous section, we neglected the noise contributions aiming at illustrating the principle of operation.
The first two terms in the right-hand side of \eqref{eqSigIdeal} are independent of the RX index $n$ and can be effectively removed at the output of the spatial DC removal block which operates across $n$ index (independently for different values of $m$) on individual snapshots. 
If we substitute from \eqref{eqDirPhase} and \eqref{eqTarPhase} into \eqref{eqSigIdealTot}, it can be shown at the output of the DC removal stage that    
\begin{subequations}\label{eqSig}
	\begin{equation}
		w_{mn}(t) = \mathrm{LPF}\{\vert A_{m}(t)\vert \cos\left (\psi_{mn}(t) + \angle A_{m}(t)\right )\},
	\end{equation}
	\begin{flalign}	
		&A_{m}(t) = 2A_{D}^{*}A_{T}^{}(t) s_{m}^{*}(t)s_{m}(t-\tau_{\rm bi}(t)),&&\\
		&\psi_{mn}(t) = \frac{2\pi}{\lambda} \left[d_{tx,m}u(t) + d_{rx,n}v(t)\right]+ \phi_{T,00}^{}(t),&&\label{eqPsi}
	\end{flalign}
\end{subequations}
where
\begin{equation}\label{equvIdeal}
	\begin{aligned}
		u(t) &\triangleq \sin\beta_{tx}(t)+\sin\alpha_{tx},\\
		v(t) &\triangleq \sin\beta_{rx}(t)+\sin\alpha_{rx}
	\end{aligned}
\end{equation}
represent the spatial frequencies associated with the DoD and DoA, respectively.
Assuming that $s_m(t),\ m=0,\ldots,M-1 $ are sample functions of a wide-sense stationary random process, it follows that
$\mathbb{E}\{s_m(t)s_m^*(t-\tau)\}=R_s(\tau)$.
In the forward-scatter region, the bistatic delays $\tau_{\mathrm{bi}}$ are confined to a narrow range, such that
$R_s(\tau_{\mathrm{bi}}(t))$ remains close to the signal power
$P_s=\mathbb{E}\{|s_m(t)|^2\}$.
As a result, the product $s_m(t)s_m^*(t-\tau_{\mathrm{bi}}(t))$ consists of a dominant DC component close to $P_s$, accompanied by zero-mean fluctuations whose spectral content is broadband relative to the Doppler band of interest (for random-like signals such as OFDM signals such fluctuations has an almost white spectrum). 
The subsequent low-pass filtering stage, whose bandwidth is selected according to the maximum expected bistatic Doppler shift,
strongly attenuates these fluctuations while preserving the DC component.
Consequently, the resulting process behaves as a nearly deterministic estimate of $R_s(\tau_{\mathrm{bi}})$ which is independent of the transmit index $m$.

Accordingly, and resorting to ULA layouts for the TX and RX arrays with $d_{\mathrm{tx},m} = md_{\mathrm{tx}}$ and 
$d_{\mathrm{rx},n} = nd_{\mathrm{rx}}$ while evaluating the expression at a snapshot $t = t_0$, we yield
\begin{equation}\label{eqWmnIdeal}
	w_{mn}(t_0) \simeq \left |\tilde{R}_s(t_0)\right | \cos\left(\frac{2\pi}{\lambda}\left(md_{\mathrm{tx}}u(t_0) + nd_{\mathrm{rx}}v(t_0)\right)  + \phi_0 \right), 
\end{equation}
where 
\begin{equation}
	\tilde{R}_s(t_0) = 2A_{D}^{*}A_{T}^{}(t)R_s(\tau_\mathrm{bi}(t_0)),
\end{equation} 
 and $\phi_0 = \phi_{T,00}(t_0) + \angle \tilde{R}_s(t_0)$. 
As is apparent from \eqref{eqWmnIdeal}, the extracted real-valued output has a similar dependence on the indices of the array element on transmit and on receive. This in turn demonstrates that a similar amplitude modulation mechanism is induced by the target across the TX and the RX arrays, depending on its instantaneous DoD and DoA, specified by spatial frequencies u and v, respectively.
Processing this two-dimensional real-valued signal $w_{mn}(t)$ via a 2D-FFT enables us to detect prospective targets in DoD-DoA domain and estimate their location. Conducting such processing over consecutive snapshots, we can generate a 3D-representation of the scene in terms of DoD-DoA-time. In particular, we get the following representation  
\begin{equation}\label{eq3dRep}
	\mathcal{V}(u,v;t) = \left|\sum_{m=0}^{M-1}\sum_{n=0}^{N-1} w_{mn}(t) e^{-\ju\frac{2\pi}{\lambda}\left(md_\mathrm{tx}u + nd_\mathrm{rx}v\right)}\right|^2.	
\end{equation}  

The instantaneous DoD–DoA information can then be mapped into two-dimensional Cartesian coordinates, given that the FSR geometry is known. As in SISO and SIMO FSR architectures, the use of real-valued input signals introduces an inherent sign ambiguity in the observable quantities, resulting in two symmetric target-location solutions on opposite sides of the baseline.

The above mathematical developments also apply to the MISO configuration by setting $N=1$ and $M>1$, demonstrating that the SIMO-FSR principle introduced in \cite{ajorloo2025forward} admits a dual formulation based on multiple baselines generated by exploiting multiple transmit (TX) elements. This formulation enables, in principle, the retrieval of the target's DoD information using a transmitter equipped with an antenna array and a single-channel receiver. Such an architecture may represent a more practical and cost-effective solution for a wide range of applications.

Nevertheless, it is worth recalling that the analysis presented in this section assumes ideal TX signals separability. This idealized scenario serves to elucidate the operating principle of MIMO-FSR and to provide a reference benchmark against which practical implementations can be assessed.

In the next two sections, we introduce two practical separation strategies based on frequency-division and time-division schemes. Although assigning distinct carrier frequencies or time slots provides a straightforward means of separating the transmitted signals, we show that these design choices significantly affect the underlying MIMO-FSR mechanism. Consequently, they must be carefully accounted for in the signal processing framework to ensure that the amplitude-based information is correctly preserved and extracted.

\section{Separation via Frequency Division Multiplexing}\label{sec:FDM}
\subsection{FDM architecture}
As a first approach to the fulfilling the separability of the transmitted waveforms on the RX side, we resort to a frequency division approach. This way, band-selective pre-filtering at each receiver can be used to easily isolate the signals before the amplitude is extracted. In such an approach, the signal emitted by the $m$th TX antenna is written as  
\begin{equation}\label{eqSigs}
	s_m(t) = \tilde{s}_{m}(t) e^{\ju2\pi m\Delta f t},\quad {\scriptstyle m=0,\ldots,M-1} 
\end{equation}
where $ \tilde{s}_{m}(t) $ is a baseband waveform with bandwidth \(\mathrm{BW}< \Delta f \). We assume that the narrowband regime also applies to the composite transmitted signal, i.e. $ M\Delta f/f_c \ll 1$. 

Thus, after band-selective pre-
filtering and amplitude extraction, \eqref{eqSigIdeal} modifies to 
\begin{align}
	y_{mn}(t)&=|A_{D}\tilde{s}_{m}(t)|^2 + |A_{T}(t)\tilde{s}_{m}(t-\tau_{\rm bi}(t))|^2\nonumber\\ 
	&+ 2 \Re\Bigl\{A_{D}^{}A_{T}^{*}(t) \tilde{s}_{m}(t)\tilde{s}_{m}^{*}(t-\tau_{\rm bi}(t))\\ 
	&\quad\qquad\qquad \cdot e^{\ju\left[\phi_{D,mn}^{}-\phi_{T,mn}^{}(t) + 2\pi m\Delta f \tau_{\rm bi}(t)\right]}\Bigr \}.\nonumber
\end{align} 
Again, at a time snapshot \( t = t_0 \), the DC-removal stage is able to cancel the first two terms. Since these terms may, in general, vary with the transmit index \(m\), the DC-removal operation can be performed across the receive antennas separately for each transmitter. Consequently, and considering the low-pass filtering effect---as explained in the previous section---we obtain at the output of the DC removal stage
\begin{equation}\label{eqwFdm}
 w_{mn}(t_0) \approx \left |\tilde{R}_{\tilde{s}}(t_0)\right |\cos(\tilde{\psi}_{mn}(t_0) + \angle \tilde{R}_{\tilde{s}}(t_0)),
\end{equation}
with
\begin{equation}
		\tilde{R}_{\tilde{s}}(t) = 2A_{D}^{*}A_{T}^{}(t) \overbrace{\mathbb{E}\{\tilde{s}_{m}^{*}(t)\tilde{s}_{m}(t-\tau_{\rm bi}(t))\}}^{R_{\tilde{s}}(\tau_{\rm bi}(t))} 
\end{equation}
and
\begin{align}\label{eqPhasefdiv}
		\tilde{\psi}_{mn}(t) &= \frac{2\pi }{\lambda} \left(  md_\mathrm{tx}u(t) + nd_\mathrm{rx}v(t) \right) + \phi_{T,00}^{}(t)\nonumber\\ 
        &- \frac{2\pi m\Delta f}{c} R_{\rm bi}(t). 
\end{align}
The first two components in \eqref{eqPhasefdiv} coincide with those in \eqref{eqPsi}. However, the third term introduces an additional dependence on the transmitter index, which is linked to the unknown bistatic range
\(
R_{\mathrm{bi}}(t_0) = R_{tx,0}(t_0) + R_{rx,0}(t_0) - B,
\)
and therefore cannot be neglected. 
This phenomenon somehow resembles the range--angle coupling encountered in conventional monostatic MIMO radar\cite{rabaste2013signal}; in MIMO-FSR, it is responsible of target angular errors if the amplitude modulation across TX elements is exploited as in \eqref{eq3dRep} to estimate the DoD.
A key advantage of the bistatic configuration, however, is that the TX and RX arrays provide two independent angle measurements, making it possible to eliminate \(R_{\mathrm{bi}}\) from the expression by rewriting it in terms of the DoA and DoD angles. To achieve this, we solve the following system of equations:
\begin{equation}
	\begin{cases}
	R_{tx,0}\sin\gamma_{tx} = R_{rx,0}\sin\gamma_{rx}\\
	R_{tx,0}\cos\gamma_{tx} + R_{rx,0}\cos\gamma_{rx} = B.
	\end{cases} 
\end{equation}
Then, it can be shown that
\begin{equation}\label{key}
	R_{\mathrm{bi}} = B\underbrace{\left [ \frac{\sin\gamma_{tx} + \sin\gamma_{rx} - \sin(\gamma_{tx}+\gamma_{rx})}{\sin(\gamma_{tx}+\gamma_{rx})} \right]}_{\triangleq g(\gamma_{tx}, \gamma_{rx})},
\end{equation}
and we can rewrite \eqref{eqwFdm} as
\begin{equation}\label{}
	w_{mn}(t_0) = \left |\tilde{R}_{\tilde{s}}(t_0)\right |\cos\left (\frac{2\pi}{\lambda}\left  ( md_{\rm tx}\tilde{u}(t_0) + nd_{\rm rx}\tilde{v}(t_0)\right) + \tilde{\phi}_0\right ),
\end{equation}
where 
\begin{equation}\label{equtilde}
\begin{cases} 
	\tilde{u}(\beta_{tx},\beta_{rx}) &=  \sin\beta_{tx}+\sin\alpha_{tx} - \frac{g(\beta_{tx}  + \alpha_{tx},\beta_{rx} + \alpha_{rx})B \Delta f}{ f_c d_\tx} \\
	\tilde{v}(\beta_{tx},\beta_{rx}) &= \sin\beta_{rx}+\sin\alpha_{rx},
\end{cases}
\end{equation}
and $\tilde{\phi}_0 $ is constant across different TX and RX antennas. Through \eqref{equtilde}, each pair of DoD-DoA parameters, i.e. $ (\beta_{tx},\beta_{rx}) $ can be mapped into a pair of $ (\tilde{u}, \tilde{v})$ and vice versa. For this case, similar to \eqref{eq3dRep}, we can obtain the 3D representation in terms of $ \tilde{u} $ and $ \tilde{v} $ as $ \mathcal{V}(\tilde{u},\tilde{v};t) $, and consequently, in terms of the DoD and DoA parameters.

\subsection{Resolution and Ambiguities}
The angular resolution of the system is conventionally defined as the -3 dB width of the main lobe in the 2D-FFT output. Expressed in terms of $\tilde{u}$ and $\tilde{v}$, the resolution is uniform across their entire domains and is given by:

\begin{equation}\label{eqResFDMuv}
	\Delta\tilde{u}_{3dB} \simeq \frac{0.891\lambda}{Md_\mathrm{tx}},\quad \Delta\tilde{v}_{3dB} \simeq \frac{0.891\lambda}{Nd_\mathrm{rx}}.
\end{equation}
However, the resolution in terms of $\beta_\mathrm{tx}$ and $\beta_\mathrm{rx}$ is position dependent. To obtain the resolution in terms of $\beta_\mathrm{tx}$ and $\beta_\mathrm{rx}$, we shall try to linearize \eqref{equtilde} around a specific point. Without affecting the derivation, we let $\alpha_\tx=\alpha_\rx=0$. Then, we have
\begin{align}
\tilde{u}(\beta_{\tx},\beta_{\rx}) &= \sin\beta_{\tx}
- K\, g(\beta_{\tx},\beta_{\rx}),\nonumber \\
\tilde{v}(\beta_{\tx},\beta_{\rx}) &= \sin\beta_{\rx},\nonumber
\end{align}
where we have defined \(K \triangleq \frac{B\Delta f}{f_c d_{\tx}}\). We linearize the mapping around the point $\tilde{u}_0(\beta_{\mathrm{tx},0},\beta_{\mathrm{rx},0})$ and $\tilde{v}_0(\beta_{\mathrm{tx},0},\beta_{\mathrm{rx},0})$ as
\begin{align}
\Delta \tilde{u} &=
\left.
\frac{\partial \tilde{u}}{\partial \beta_{\tx}}
\right|_{0}
\Delta \beta_{\tx}
+
\left.
\frac{\partial \tilde{u}}{\partial \beta_{\rx}}
\right|_{0}
\Delta \beta_{\rx}, \\
\Delta \tilde{v} &=
\left.
\frac{\partial \tilde{v}}{\partial \beta_{\rx}}
\right|_{0}
\Delta \beta_{\rx}.
\end{align}
where ``$|_{0}$'' stands for the evaluation of the derivative at $(\beta_{\tx,0},\beta_{\rx,0})$. Evaluating the derivatives gives
\begin{align}
\Delta \tilde{u} &=
\left(
\cos\beta_{\tx,0}
- K \frac{\partial g}{\partial \beta_{\tx}}
\bigg|_{0}
\right)\Delta \beta_{\tx}
-
K \frac{\partial g}{\partial \beta_{\rx}}
\bigg|_{0}
\Delta \beta_{\rx}, \\
\Delta \tilde{v} &=
\cos\beta_{\rx,0}\,\Delta \beta_{\rx}.
\end{align}
For the DoD resolution (along $\beta_\tx$), $\beta_\rx$ is held fixed
(\(\Delta \beta_{\rx}=0\)), yielding
\begin{align}\label{eq28}
\Delta \tilde{u}
=
\left(
\cos\beta_{\tx,0}
- K \frac{\partial g}{\partial \beta_{\tx}}
\bigg|_{0}
\right)\Delta \beta_{\tx}.
\end{align}
For the receive-angle resolution, set \(\Delta \beta_{\tx}=0\), giving
\begin{align}
\Delta \tilde{u}
&=
- K \frac{\partial g}{\partial \beta_{\rx}}
\bigg|_{0}
\Delta \beta_{\rx}, \\
\Delta \tilde{v}
&=
\cos\beta_{\rx,0}\,\Delta \beta_{\rx}.
\end{align}
Indeed, a variation in \(\beta_{\rx}\) affects both FFT coordinates. We can approximate
\(-3\) dB contour in the \((\tilde{u},\tilde{v})\) domain by an ellipse
\begin{align}
\frac{\tilde{u}^2}{\Delta \tilde{u}^2}
+
\frac{\tilde{v}^2}{\Delta \tilde{v}^2}
=1.
\end{align}
Substituting the linearized relations yields
\begin{align}\label{eq32}
\left[
\left(
\frac{
K \frac{\partial g}{\partial \beta_{\rx}}\big|_{0}
}{\Delta \tilde{u}}
\right)^2
+
\left(
\frac{
\cos\beta_{\rx,0}
}{\Delta \tilde{v}}
\right)^2
\right]
\Delta \beta_{\rx}^2
=1.
\end{align}
Finally, rearranging \eqref{eq28} and \eqref{eq32}, the DoA and DoD resolution are obtained as
\begin{align}
\Delta \beta_{\tx,3dB}
&=
\frac{\Delta \tilde{u}_{3dB}}
{\left|
\cos\beta_{\tx,0}
- K \frac{\partial g}{\partial \beta_{\tx}}
\bigg|_{0}
\right|}, \label{eqResFdmDod}\\
\Delta \beta_{\rx,3dB}
&=
\left[
\left(
\frac{
K \frac{\partial g}{\partial \beta_{\rx}}\big|_{0}
}{\Delta \tilde{u}_{3dB}}
\right)^2
+
\left(
\frac{
\cos\beta_{\rx,0}
}{\Delta \tilde{v}_{3dB}}
\right)^2
\right]^{-1/2}.\label{eqResFdmDoa}
\end{align}
where $\Delta \tilde{u}_{3dB}$ and $\Delta \tilde{v}_{3dB}$ are given by \eqref{eqResFDMuv}.


Letting $K=0$ in the above equations (i.e., no correction of the range–angle coupling), we end up with the resolution for the ideal case. With correction applied, the -3 dB contour of the output is deformed, since the remapping induces an intrinsic dependence between the two observables. As a consequence, the DoA resolution is somewhat improved with respect to the ideal case, irrespective of the sign of the partial derivative. However, the DoD resolution is not symmetrically identical in either side of the baseline. To illustrate this, let us consider the typical case of no tilting $(\alpha_\mathrm{tx} = \alpha_\mathrm{rx} = 0)$. When the targets lies above the baseline, i.e., $\beta_\mathrm{tx},\beta_\mathrm{rx} > 0$ (corresponding to the first quadrant of the DoD–DoA map), $\frac{\partial g}{\partial \beta_\mathrm{rx}}$ is positive which causes the DoD resolution to degrade with respect to the ideal case. Conversely, when the target is below the baseline, i.e., $\beta_\mathrm{tx},\beta_\mathrm{rx} < 0$ (corresponding to the third quadrant of the DoD–DoA map), the partial derivative is negative, which causes the DoD resolution to improve with respect to the ideal case. 
 
The correction also affects the ambiguity structure. For a target located at $(\beta_{\mathrm{tx},0},\beta_{\mathrm{rx},0})$, corresponding to $(\tilde{u}_0,\tilde{v}_0) = (\tilde{u}(\beta_{\mathrm{tx},0},\beta_{\mathrm{rx},0}),\tilde{v}(\beta_{\mathrm{tx},0},\beta_{\mathrm{rx},0}))$, an intrinsic ambiguous peak appears at $(-\tilde{u}_0,-\tilde{v}_0)$ due to the real-valued nature of the input signal. Considering the typical no-tilting case $(\alpha_\mathrm{tx} = \alpha_\mathrm{rx} = 0)$, such symmetry in $\tilde{u}-\tilde{v}$ plane---unlike the ideal case---does not translate into a symmetric peak at $(-\beta_{\mathrm{tx},0},-\beta_{\mathrm{rx},0})$. 

In general, the set of all possible peaks in the $\tilde{u}-\tilde{v}$ plane (irrespective of the field of view (FOV)), is            
\begin{equation}
 	\left \{\pm(\tilde{u}_0,\tilde{v}_0) + \left(k\frac{\lambda}{d_\mathrm{rx}}, k'\frac{\lambda}{d_\mathrm{rx}}\right) \Big| k,k'\in \mathbb{Z} \right\}
\end{equation}
Among these peaks, ideally, only $\pm (\tilde{u}_0,\tilde{v}_0)$ shall lead to valid pairs of $(\beta_\tx,\beta_\rx)$ in the FOV and the remaining peaks should lie outside the FOV.    
Let the FOV be defined as $[-\beta_\mathrm{tx,max},\beta_\mathrm{tx,max}]$ and $[-\beta_\mathrm{rx,max},\beta_\mathrm{rx,max}]$ in the DoD and DoA domains, respectively, and assume $\alpha_\mathrm{tx} = \alpha_\mathrm{rx} = 0$. The condition for avoiding ambiguity due to insufficient inter-element spacing is then
\begin{equation}
	\begin{aligned}
	&-\sin\beta_\mathrm{tx,max} - \frac{\alpha_\mathrm{FDM} B}{d_\tx} g(-\beta_\mathrm{tx,max},  -\beta_\mathrm{rx,max}) + \frac{\lambda}{d_\tx}> 1\\
	&-\sin\beta_\mathrm{rx,max} + \frac{\lambda}{d_\rx}> 1
\end{aligned}
\end{equation}
where $\alpha_\mathrm{FDM}\triangleq\frac{\Delta f}{f_c}$ is the relative FDM separation. Since
$g(-\beta_\mathrm{tx},-\beta_\mathrm{rx}) = g(\beta_\mathrm{tx},\beta_\mathrm{rx})$, this yields
\begin{equation}\label{eqAmbFDM}
	\begin{aligned}
	d_\tx &< \frac{\lambda-\alpha_\mathrm{FDM} R_\mathrm{bi,max}}{1 + \sin\beta_\mathrm{tx,max}}\\
d_\rx &< \frac{\lambda}{1 + \sin\beta_\mathrm{rx,max}}.	
\end{aligned} 
\end{equation}
where $R_\mathrm{bi,max} = B g(\beta_\mathrm{tx,max}, \beta_\mathrm{rx,max})$ is the maximum bistatic range in the FOV. The condition on $d_\rx$ is the same as in the ideal case and is unaffected by the correction ($d_\rx<\lambda/2$ in case of a full FOV).
In contrast, the constraint on $d_\tx$ is highly dependent on the value of $\alpha_\mathrm{FDM}$. 
As $\alpha_\mathrm{FDM}$ increases, this constraint becomes
progressively more restrictive; in particular, for
$\alpha_\mathrm{FDM}>\lambda/R_\mathrm{bi,max}$, no choice of
$d_\mathrm{tx}$ can completely eliminate spatial ambiguities within the prescribed FOV.

The coonditions in \eqref{eqAmbFDM} guarantees that no grating lobe generated by any target within the visible region falls inside the prescribed FOV. If, however, it is known \emph{a priori} that targets can only exist within the prescribed angular sector, a less restrictive design criterion can be adopted. In this case, it is sufficient that the spatial-frequency interval associated with the FOV does not overlap with its adjacent periodic replicas, yielding the sector-limited condition
\begin{equation}\label{eqAmbFDM2}
	\begin{aligned}
	d_\tx &< \frac{\lambda-\alpha_\mathrm{FDM} R_\mathrm{bi,max}}{2 \sin\beta_\mathrm{tx,max}}\\
d_\rx &< \frac{\lambda}{2 \sin\beta_\mathrm{rx,max}}.	
\end{aligned} 
\end{equation}

\subsection{Trade-offs and Limitations}
In this subsection, we discuss the practical implications of the FDM approach. Indeed, the practical implementation of FDM introduces several critical trade-offs, particularly concerning spatial distortions and hardware configurations. These include:



\begin{itemize}
    \item \textbf{Implementation of corrective terms:} In the FDM approach, the availability of both TX and RX arrays allows the corrective term to be incorporated smoothly into the MIMO processing framework without introducing additional unknown parameters as it is specified by the same unknowns of the system, namely DoA and DoD along with the known system parameters (baseline $B$ and frequency separation $\Delta f$). In contrast, as discussed in the next section, the TDM approach needs the introduction of an additional unknown parameter to enable the corresponding correction.

    \item \textbf{MISO limitations:} A major drawback of FDM is its incompatibility with pure MISO configurations. Without an antenna array at the receiver to provide spatial diversity, the corrective term cannot be expressed as a function of the DoA. Consequently, FDM cannot support accurate DoD-time processing for cost-effective, single-antenna receivers.
    
    \item \textbf{Output map distortions and error scaling:} The necessary FDM corrective term scales with the product $B\Delta f$. For medium-to-large baselines or large frequency separations, failing to apply this correction results in significant DoD errors. However, applying the correction introduces nonlinear map distortions, including direction-dependent resolution variations, target peak deformations, and ambiguities within the field of view, which can notably degrade detection and estimation performance.
    
    \item \textbf{Range applicability:} Because these distortions worsen as $B\Delta f$ increases---and significantly reducing $\Delta f$ is often technologically restricted---FDM becomes particularly challenging for medium- and long-range applications.
    
    \item \textbf{Tracking update rates:} Conversely, FDM offers an advantage in target tracking update rates and robustness against Doppler migration; because all transmitters illuminate the scene simultaneously, the angular map can be updated almost instantaneously at the temporal snapshot level.
\end{itemize}
     
\section{Separation via Time Division Multiplexing}\label{sec:TDM}
\subsection{TDM architecture}
As the second approach to the separability issue, we opt to use a time division strategy. In particular, we consider a case where different transmitters operate in distinct time intervals, ensuring that the signal received at the RX side at any instant originates from only one transmitter. Letting $\Delta T$ denote the temporal spacing between consecutive transmissions, we express the transmitted waveforms as
\begin{equation}\label{eqWavTdm}
	s_m(t) = \tilde{s}_m(t-m\Delta T)\cdot{\rm rect}_{\Delta T} \left(t-m\Delta T\right),\: {\scriptstyle m=0,\ldots,M-1}.   
\end{equation}
where $ {\rm rect}_{T}(t) \triangleq u(t) - u(t-T) $ is the pulse function of duration $ T $.   
Evaluating $x_n(t)$, as given by \eqref{eq_xn}, at 
\begin{equation}
	t_m = t_0 + m\Delta T,\quad 0<t_0<\Delta T,
\end{equation}
gives: 
\begin{align}\label{eqxTdm}
	x_{n}(t_m) &=  A_{D}^{} s_m(t_m)  e^{j\phi_{D,mn}^{}}\\ 
	&+ A_{T}^{}(t_m) s_m(t_m-\tau_{\rm bi}(t_m)) e^{j\phi_{T,mn}^{}(t_m)} + \zeta_n(t_m).\nonumber
\end{align}   
Next, we use \eqref{eqWavTdm} in \eqref{eqxTdm} and express the received signals as indexed by both $m$ and $n$. it follows that 
\begin{align}
	x_{n}(t_m) =& x_{mn}(t_0) =  A_{D}^{} \tilde{s}_m(t_0)  e^{j\phi_{D,mn}^{}}\\ 
	&+ A_{T}^{}(t_0) \tilde{s}_m(t_0-\tau_{\rm bi}(t_0)) e^{j\phi_{T,mn}^{}(t_m)} + \zeta_n(t_m),\nonumber
\end{align}   
where we used the approximations $A_T(t_m)\approx A_T(t_0)$ and $\tilde{s}_m(t_0-\tau_{\rm bi}(t_m)) \approx \tilde{s}_m(t_0-\tau_{\rm bi}(t_0))$. For every $x_{mn}(t), m=0,\dots,M-1, n=0,\dots,N-1 $, the same processing chain as depicted in Fig.\ref{FigProcChain1Channel}, is applied which starts with squaring the signal. The squared magnitude expression is given by
\begin{align}\label{eqSigTdm}
y_{mn}(t_0)
=&\,|x_{mn}(t_0)|^2 \\
=&\,|A_{D}\tilde{s}_{m}(t_0)|^2
+|A_{T}(t_0)\tilde{s}_{m}(t_0-\tau_{\rm bi}(t_0))|^2
\nonumber\\
&+2\Re\Bigl\{
A_{D}A_{T}^{*}(t_0)\,
\tilde{s}_{m}(t_0)
\tilde{s}_{m}^{*}(t_0-\tau_{\rm bi}(t_0))
\nonumber\\
&\hspace{1.9cm}\times
\mathrm{e}^{j\left[\phi_{D,mn}-\phi_{T,mn}(t_m)\right]}
\Bigr\}.
\end{align}
For a sufficient number of RX antennas $N$, applying DC removal across the receivers, separately for each transmitter, effectively removes the first two terms in \eqref{eqSigTdm}. Taking into account also the impact of low-pass filtering, we therefore obtain
\begin{equation}\label{eqwTdm}
	w_{mn}(t_0) = |\tilde{R}_{\tilde{s}}(t_0)|\cos(\tilde{\psi}_{mn}(t_0) + \angle \tilde{R}_{\tilde{s}}(t_0)),
\end{equation}
with
\begin{equation}
	\tilde{R}_{\tilde{s}}(t) = 2A_{D}^{*}A_{T}^{}(t) R_{\tilde{s}}(\tau_{\rm bi}(t)) 
\end{equation}
and
\begin{equation}\label{eqPhaseTdiv}
	\tilde{\psi}_{mn}(t_0) = -\frac{2\pi}{\lambda}\!\left(R_{\mathrm{tx},m}(t_m) 
	+ R_{\mathrm{rx},n}(t_m) - B \right) - \phi_{D,mn}. 
\end{equation}      
Assuming that the target’s velocity, DoA, and DoD remain approximately constant over the full transmission cycle of duration $M\Delta T$, we have
\begin{align}
	R_{\mathrm{tx},m}(t_m) 
	&\simeq R_{\mathrm{tx},0}(t_m) - m d_\mathrm{tx} \sin(\beta_{\mathrm{tx}}) \\
	&= R_{\mathrm{tx},0}(t_0) + m \Delta T \, \dot{R}_{\mathrm{tx}}
	- m d_\mathrm{tx} \sin(\beta_{\mathrm{tx}})\nonumber
\end{align}
\begin{align}
	R_{\mathrm{rx},n}(t_m) 
	&\simeq R_{\mathrm{rx},0}(t_m) - n d_\mathrm{rx} \sin(\beta_{\mathrm{rx}}) \\
	&= R_{\mathrm{rx},0}(t_0) + m \Delta T \, \dot{R}_{\mathrm{rx}}
	- n d_\mathrm{rx} \sin(\beta_{\mathrm{rx}}),\nonumber
\end{align}
substituting these equations into \eqref{eqPhaseTdiv}, gives
\begin{equation}\label{eqPhaseTdiv1}
	\tilde{\psi}_{mn}(t_0) = \phi_{T,00}(t_0) + \frac{2\pi}{\lambda}\!\left(md_\mathrm{tx}\tilde{u} 
	+ nd_\mathrm{tx}\tilde{v} \right).
\end{equation}
where 
\begin{equation}\label{eqSpFreqsTdm}
	\begin{aligned}
	\tilde{u}&=\sin \alpha_{tx} + \sin \beta_{tx} - \frac{\Delta T\dot{R}_\mathrm{bi}}{d_\mathrm{tx}}\\ 
\tilde{v}&=\sin \alpha_{rx} + \sin \beta_{rx} 
\end{aligned}
\end{equation} 
and $\dot{R}_\mathrm{bi} = \dot{R}_\mathrm{tx} + \dot{R}_\mathrm{tx}$ is the bistatic range rate, i.e., the bistatic velocity. 
As in the FDM case, the adoption of TDM inherently alters the dependence of the signal amplitude modulation on the target kinematic parameters relative to the ideal case in \eqref{equvIdeal}. In particular, the distortion introduced by the TDM scheme is determined by the target bistatic velocity, the temporal interval between consecutive transmissions, and the spacing between the TX array elements.

If $\dot{R}_{\mathrm{bi}}$ were known, the spatial frequency correction in \eqref{eqSpFreqsTdm} could be applied directly, and a standard 2D-FFT would yield the corrected DoD--DoA representation.  
In practice, however, this parameter is unknown at the receiver. 
However, unlike the FDM case, the unknown quantity affecting the signal amplitude modulation across the TX elements is directly measurable from the received data on individual receivers.
Indeed, since it is proportional to the bistatic Doppler shift, the bistatic range rate can be estimated independently using conventional SISO time-domain processing applied to the temporal samples collected by a single TX--RX pair.

Alternatively, we can integrate the Doppler processing into our spatial processing by including the temporal samples gathered at all intervals. To this end, let us consider the sampling frequency such that $\Delta T = L T_s$. That is, we collect $L$ temporal samples at each transmission interval. We define $t_{ml} = m\Delta T + lT_s $ as the time of $l$th sample at $m$th interval. This way, $\tilde{\psi}$ in \eqref{eqPhaseTdiv} will be subject to three indices $m$, $n$, and $l$:
\begin{equation}\label{eqPhaseTdm2}
		\tilde{\psi}_{mnl}(t_0) = -\frac{2\pi}{\lambda}\!\left(R_{\mathrm{tx},m}(t_{ml}) 
+ R_{\mathrm{rx},n}(t_{ml}) - B \right) - \phi_{D,mn}. 
\end{equation}
where
\begin{align}
	R_{\mathrm{tx},m}(t_{ml}) 
	\simeq R_{\mathrm{tx},0}(t_0) + (m\Delta T + lT_s) \dot{R}_{\mathrm{tx}} - m d_\mathrm{tx} \sin(\beta_{\mathrm{tx}}) \\
	R_{\mathrm{rx},m}(t_{ml}) 
	\simeq R_{\mathrm{rx},0}(t_0) + (m\Delta T + lT_s) \dot{R}_{\mathrm{tx}} - n d_\mathrm{rx} \sin(\beta_{\mathrm{tx}}) \nonumber
\end{align}
substituting into \eqref{eqPhaseTdm2} yields:
\begin{equation}\label{eqPhaseTdiv2}
	\tilde{\psi}_{mnl}(t_0) = \phi_{T,00}(t_0) + \frac{2\pi}{\lambda}\!\left(md_\mathrm{tx}\tilde{u} 
	+ nd_\mathrm{tx}\tilde{v} + l \tilde{w} \right).
\end{equation}
where 
\begin{equation}\label{equvwtilde}
	\begin{cases} 
		\tilde{u}(\beta_{tx},\beta_{rx}, \dot{R}_\mathrm{bi}) &=  \sin\beta_{tx}+\sin\alpha_{tx} - \frac{\Delta T\dot{R}_\mathrm{bi}}{d_\mathrm{tx}} \\
		\tilde{v}(\beta_{tx},\beta_{rx},\dot{R}_\mathrm{bi}) &= \sin\beta_{rx}+\sin\alpha_{rx}\\
		\tilde{w}(\beta_{tx},\beta_{rx}, \dot{R}_\mathrm{bi}) &= T_s \dot{R}_\mathrm{bi}
	\end{cases}
\end{equation}
Through \eqref{equvwtilde}, each triple of the desired parameters, namely $(\beta_\mathrm{tx},\beta_\mathrm{rx},\dot{R}_\mathrm{bi})$   can be mapped into a triple of $ (\tilde{u}, \tilde{v},\tilde{w})$ and vice versa. Therefore, through a 3D-FFT over the 3D real-valued data samples (TX,RX,time), one can generate a representation of the scene and prospective targets in terms of $\tilde{u}$, $\tilde{v}$, and $\tilde{w}$ and consequently in terms of $\beta_\mathrm{tx}$, $\beta_\mathrm{rx}$, and $\dot{R}_\mathrm{bi}$. Repeating the same processing over consecutive batches of the received signals of length $M\Delta T$, we will be able to acquire the same 3D representations over time, to track the evolution of the target scene.    

\subsection{Resolution and Ambiguities} To obtain expressions for angular resolution, we shall follow a similar approach as in the FDM case and linearize \eqref{eqSpFreqsTdm} around ($\beta_{\mathrm{tx},0},\beta_{\mathrm{rx},0}$). Considering the no-tilt case ($\alpha_\tx=\alpha_\rx=0$), we express $ \dot{R}_\mathrm{bi} = \mathbf{v} \cdot (\hat{\mathbf{s}}_\mathrm{tx} + \hat{\mathbf{s}}_\mathrm{rx})$ where $ \bv$ is the velocity vector, and 
\begin{equation}
\hat{\mathbf{s}}_\mathrm{tx}  =
\begin{bmatrix}
\cos\beta_\tx \\ \sin\beta_\tx
\end{bmatrix},\quad\hat{\mathbf{s}}_\mathrm{rx}=
\begin{bmatrix}
	-\cos\beta_\rx \\ \sin\beta_\rx
\end{bmatrix}
\end{equation}
denote the unit vector from the TX to the target and from the RX to the target, respectively. Linearizing gives
\begin{equation}
	\begin{aligned}
		\Delta\tilde{u} &= \Delta\beta_\tx \cos\beta_{\tx,0} \\
        &- \frac{\Delta T}{d_\tx} \left ( \bv\cdot\ \frac{\partial\hat{\mathbf{s}}_\mathrm{tx}}{\partial \beta_\mathrm{tx}}\bigg|_{0} \Delta \beta_\tx  + \bv\cdot\ \frac{\partial\hat{\mathbf{s}}_\mathrm{rx}}{\partial \beta_\mathrm{rx}}\bigg|_{0} \Delta\beta_\rx \right)\\
		\Delta\tilde{v} &= \Delta\beta_\rx \cos\beta_{\rx,0}
	\end{aligned}
\end{equation} 
Adopting the same approach as in the FDM case, we obtain
\begin{align}
	\Delta\beta_\tx &= \frac{\Delta\tilde{u}}{\cos\beta_{\tx,0} - \frac{\Delta T}{d_\tx} \bv\cdot\ \frac{\partial\hat{\mathbf{s}}_\mathrm{tx}}{\partial \beta_\mathrm{tx}}\big|_{0}},\label{eqDeltabetaTxTdm}\\
	\Delta\beta_\rx &= \left[\left(\frac{\cos\beta_{\rx,0}}{\Delta\tilde{v}}\right)^2 + \left(\frac{\frac{\Delta T}{d_\tx} \frac{\partial\hat{\mathbf{s}}_\mathrm{tx}}{\partial \beta_\mathrm{tx}}\big|_{0}}{\Delta \tilde{u}}
\right)^2 \right]^{-1/2}
\end{align}
where $\Delta\tilde{u}$ and $\Delta\tilde{v}$ are given by \eqref{eqResFDMuv} and
\begin{equation}
	\frac{\partial\hat{\mathbf{s}}_\mathrm{tx}}{\partial \beta_\mathrm{tx}}  =
	\begin{bmatrix}
		-\sin\beta_\tx \\ \cos\beta_\tx
	\end{bmatrix},\quad\frac{\partial\hat{\mathbf{s}}_\mathrm{rx}}{\partial \beta_\mathrm{rx}}=
	\begin{bmatrix}
		\sin\beta_\rx \\ \cos\beta_\rx
	\end{bmatrix}.
\end{equation}

For the special case of a target crossing the baseline perpendicularly, that is $ \bv = [0,V]^T$ where $V$ denotes the target speed, equation \eqref{eqDeltabetaTxTdm} simplifies to
\begin{align}
	\Delta\beta_\tx &= \frac{\Delta\tilde{u}}{\cos\beta_{\tx,0} \left (1- \frac{\Delta T V}{d_\tx}\right)},\\
    \Delta\beta_\rx &=  \left[\left(\frac{\cos\beta_{\rx,0}}{\Delta\tilde{v}}\right)^2 + \left(\frac{\Delta T V \cos\beta_{\tx,0}}{d_\tx \Delta \tilde{u}}
\right)^2 \right]^{-1/2}
\end{align}
In this scenario, the DoA resolution somewhat improves with respect to the ideal case (corresponding to \(\Delta T = 0\)) irrespective of the velocity sign. For the DoD resolution, however, when the target moves along the negative \(y\)-axis ($V < 0$), the angular resolution improves relative to the ideal case. Conversely, if the target moves along the positive \(y\)-axis ($V > 0$), the resolution degrades.
The parameter governing this modification is \(\alpha_\mathrm{TDM} \triangleq \frac{\Delta T V}{d_\mathrm{tx}},\)
which represents the ratio between the distance traveled by the target during \(\Delta T\) and the transmit inter-element spacing. For small values of \(\alpha_\mathrm{TDM}\), the deviation from the ideal resolution remains negligible. 

Given an FOV of $[-\beta_\mathrm{tx,max},\beta_\mathrm{tx,max}]$ and $[-\beta_\mathrm{rx,max},\beta_\mathrm{rx,max}]$ in the DoD and DoA domains, respectively, for avoiding ambiguity we shall have
\begin{equation}
	\begin{aligned}
	&-\sin\beta_\mathrm{tx,max} - \frac{\Delta T \dot{R}_\mathrm{bi,max}}{d_\tx} + \frac{\lambda}{d_\tx} > 1\\
	&-\sin\beta_\mathrm{rx,max} + \frac{\lambda}{d_\rx}> 1
	\end{aligned}
\end{equation}
which gives
\begin{equation}
	\begin{aligned}
		d_\tx &< \frac{\lambda-\Delta T \dot{R}_\mathrm{bi,max}}{1 + \sin\beta_\mathrm{tx,max}}\\
		d_\rx &< \frac{\lambda}{1 + \sin\beta_\mathrm{rx,max}},	
	\end{aligned} 
\end{equation}
and the less conservative sector-limited condition (for a physically-restricted FOV) is given by
\begin{equation}\label{AmbTDM2}
	\begin{aligned}
		d_\tx &< \frac{\lambda-\Delta T \dot{R}_\mathrm{bi,max}}{2\sin\beta_\mathrm{tx,max}}\\
		d_\rx &< \frac{\lambda}{2\sin\beta_\mathrm{rx,max}}.	
	\end{aligned} 
\end{equation}
Again, the ambiguity in DoA is the same as the ideal case and can be avoided by letting $d_\rx<\lambda/2$ in the worst case. In the DoD domain, however, the ambiguity is governed by $\Delta T \dot{R}_\mathrm{bi,max}$; the larger this value, the more ambiguities we have.  

An interesting feature enabled by the 3D processing is the ability to resolve the sign ambiguity by exploiting the temporal evolution of the response. Consider the triplet $(\beta_{\mathrm{tx},0}, \beta_{\mathrm{rx},0}, \dot{R}_0)$ as the true target parameters. The corresponding ambiguous response appears at \((-\beta_{\mathrm{tx},0}, -\beta_{\mathrm{rx},0}, -\dot{R}_0)\). If $\dot{R}_0 > 0$, meaning that the target is moving away from the baseline, the absolute values $|\beta_\mathrm{tx}|$ and $|\beta_\mathrm{rx}|$ increase over time. However, this behavior is inconsistent with the symmetric response $(-\beta_{\mathrm{tx},0}, -\beta_{\mathrm{rx},0}, -\dot{R}_0)$, for which $\dot{R} < 0$, implying that the target would be moving toward the baseline and therefore that $|\beta_\mathrm{tx}|$ and $|\beta_\mathrm{rx}|$ should decrease over time. Conversely, if $\dot{R}_0 < 0$ the $\vert \beta_\tx \vert$ and $\vert \beta_\rx \vert$ should decrease over time which contradicts the behavior implied by the symmetric solution.           

\subsection{Trade-offs and Limitations}

Because the distortions in TDM are driven by target velocity rather than the spatial baselines used in FDM, its principle of operation reveals some distinct practical advantages and compromises compared to frequency-domain separation:

\begin{itemize}
    \item \textbf{Implementation of corrective terms:} Unlike FDM, the TDM corrective term relies on the target's bistatic velocity, $\dot{R}_{b}$. Utilizing it requires either an independent bistatic Doppler estimate (such as from SISO time-domain processing) or joint 3-D processing over DoD, DoA, and Doppler, which inherently increases the computational burden.
    
    \item \textbf{MISO Applicability:} TDM is highly suitable for MISO processing because its velocity-dependent correction does not require an RX array. A single-antenna receiver can successfully collect temporal samples across individual transmit slots to perform 2D processing in terms of DoD and bistatic Doppler. This enables the opportunistic use of a single-antenna receiver for cost-effective angular processing when a multi-antenna transmitter is available. 
  
    \item \textbf{Error scaling and map integrity:} The angular error in TDM is governed by the product $\dot{R}_{b}\Delta T$. For targets that are not excessively fast, and with a sufficiently short slot duration ($\Delta T$), the DoD error can remain negligible even if no explicit correction is applied. Furthermore, when the correction is applied, it does not significantly distort the map, making it highly advantageous compared to the structural distortions seen in FDM.

    \item \textbf{The role of slot duration ($\Delta T$):} The performance of a TDM system heavily relies on selecting an appropriate slot duration. Shorter slots are favorable because they reduce uncorrected DoD errors and nonlinear map distortions, while also improving robustness against target migration effects that occur when a target moves during the transmission frame.
    
    \item \textbf{Tracking update rates:} A shorter $\Delta T$ is also vital for tracking applications. Because TDM requires the acquisition of the entire transmission frame (all slots from all transmitters) before full MIMO information is available, the tracking update rate is directly tied to the total frame duration.
    
    \item \textbf{System design compromises:} Despite the benefits of short slots, practical TDM design must balance technological processing constraints and the potential need for a larger $\Delta T$ to ensure sufficient Doppler resolution when prior velocity information is unavailable. Nonetheless, TDM generally serves as the more robust solution for typical medium- and long-range scenarios.
\end{itemize}

\section{Numerical Simulations} \label{sec:Simulations} 
This section assesses the proposed MIMO forward-scatter processing approach in two representative simulated scenarios. The parameters common to both cases are listed in Table~\ref{tab:common_settings}. In particular, we consider a system operating at 2.4 GHz, with the transmitting and receiving arrays consisting of $10$ elements each, both with half-wavelength spacing. Two multiplexing strategies are adopted: FDM with adjacent transmitters separated by $\Delta f=1$ MHz, and TDM with an inter-transmission delay of $\Delta T=2$ ms. In all simulations, the transmitted waveform is a monochromatic sinusoid. Two different target and geometry configurations are considered. The first one is a short-range scenario with a $40$ m baseline and a human-like target  (see details in  Table~\ref{tab:scenario_human}). The second one is a long-range scenario with a $2000$ m baseline and a target resembling a small UAV (see Table~\ref{tab:scenario_uav}). In both cases, the target moves perpendicularly to the baseline and at reference time t=0 s it crosses the baseline at its midpoint, so that true DoD and DoA remain equal throughout the target path, i.e., the target peak is expected to move along the bisector of the first and third quadrants across time. 

\begin{table}[!htp]
	\caption{General settings common to both simulated scenarios}
	\label{tab:common_settings}
	\centering
	\begin{tabular}{cc}
		\hline
		\textbf{Setting} & \textbf{Value} \\ \hline
		Carrier frequency                   & $f_c = 2.4$ GHz \\ 
		Number of TX antennas               & $N_t = 10$ \\ 
		Number of RX antennas               & $N_r = 10$ \\ 
		Inter-antenna spacing               & $d_\tx=d_\rx = \lambda/2$ \\
		FDM frequency separation                & $\Delta f = 1$ MHz \\
        TDM time shift                      & $\Delta T = 2$ ms \\ 
		TX array tilt angle                & $\alpha_{tx} = 0$ \\  
		RX array tilt angle                & $\alpha_{rx} = 0$ \\  
		Waveform                            & Sinusoid \\
        Direct signal power to noise ratio        & $\mathrm{DNR} = 35 \mathrm {dB}$\\
        Initial sampling rate               & $f_s = 10 $ MHz\\
        \hline
	\end{tabular}
\end{table}

\begin{table}[!htp]
	\caption{Scenario 1: human target with short baseline}
	\label{tab:scenario_human}
	\centering
	\begin{tabular}{cc}
		\hline
		\textbf{Setting} & \textbf{Value} \\ \hline
		Baseline length                     & $B = 40$ m \\ 
		TX location                         & $(0,0)$ \\ 
		RX location                         & $(40,0)$ \\ 
		Target                              & Human \\
		Target dimensions                   & $L_h = 0.2$ m,\ $L_v = 1.7$ m \\
		Target velocity vector              & $\mathbf{v}_{\text{target}} = (0,-8\,\mathrm{km/h})$ \\
		Crossing point                      & $(20,0)$ \\ \hline
	\end{tabular}
\end{table}

\begin{table}[!htp]
	\caption{Scenario 2: small UAV with long baseline}
	\label{tab:scenario_uav}
	\centering
	\begin{tabular}{cc}
		\hline
		\textbf{Setting} & \textbf{Value} \\ \hline
		Baseline length                     & $B = 2000$ m \\ 
		TX location                         & $(0,0)$ \\ 
		RX location                         & $(2000,0)$ \\ 
		Target                              & Small UAV \\
		Target dimensions                   & $L_h = 0.3$ m,\ $L_v = 0.4$ m \\
		Target velocity vector              & $\mathbf{v}_{\text{target}} = (0,-40\,\mathrm{km/h})$ \\
		Crossing point                      & $(1000,0)$ \\ \hline
	\end{tabular}
\end{table}

\begin{figure*}[!ht]
	\centering
\includegraphics[width =\textwidth]{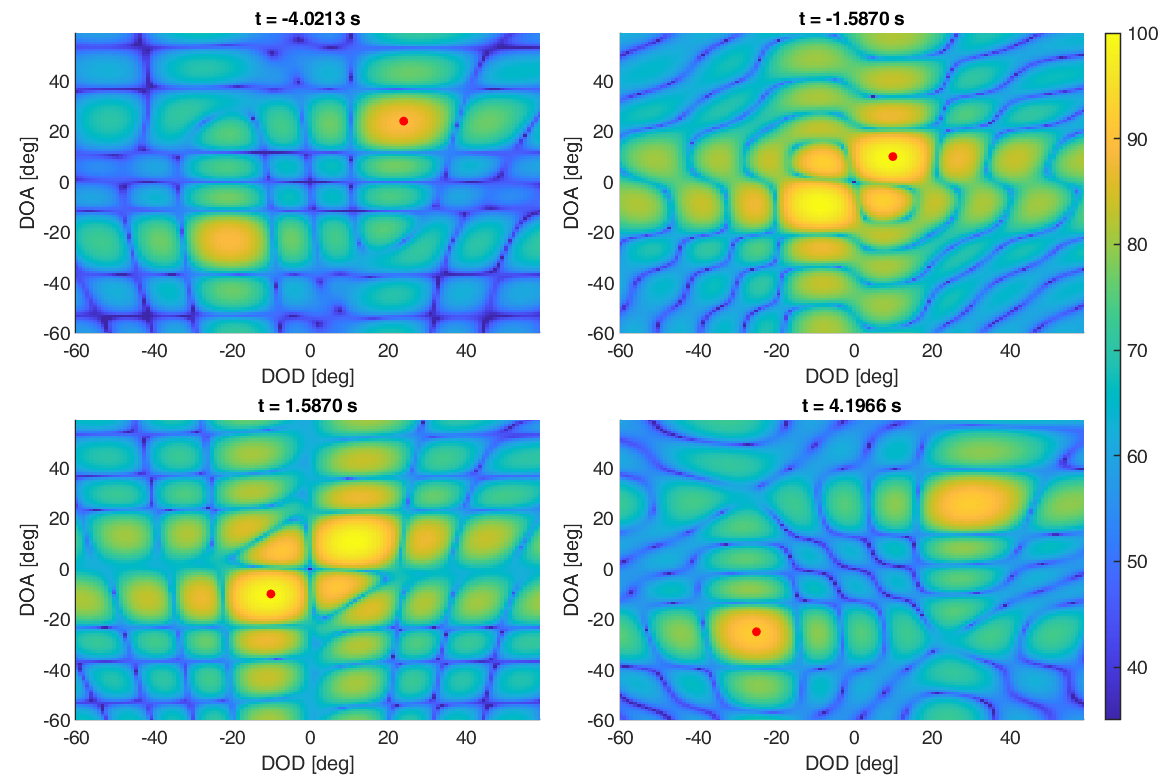}
	\caption{Four frames of DoD-DoA maps over time obtained using MIMO approach for the short baseline scenario using FDM. The red dot shows the true target location}
    \label{fig:fdm_tracking_short}
\end{figure*}  

\subsection{FDM}
Fig.~\ref{fig:fdm_tracking_short} shows four representative DoD--DoA maps obtained with the FDM approach at four snapshots for the short-baseline scenario after applying the correction described in Section \ref{sec:FDM}. At each selected frame, the true target position in the DoD--DoA plane is indicated with a red circle. As can be observed, the MIMO approach is able to localize the target consistently over time, producing focused peaks without noticeable spreading or anomalous coupling between the two angular dimensions. As expected, the true peak comes with the sign ambiguous copy due to the amplitude-based FSR principle. Despite this well known effect, in this short-range configuration, the FDM scheme provides accurate and well-behaved angular maps. 

\begin{figure*}[!ht]
    \centering

    \begin{subfigure}{0.48\textwidth}
        \centering
        \includegraphics[width=\linewidth]{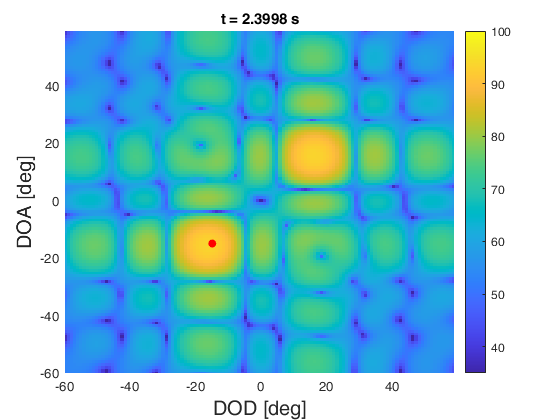}
        \caption{Short baseline without correction.}
        \label{fig:fdm_short_nocorrection}
    \end{subfigure}
    \hfill
    \begin{subfigure}{0.48\textwidth}
        \centering
        \includegraphics[width=\linewidth]{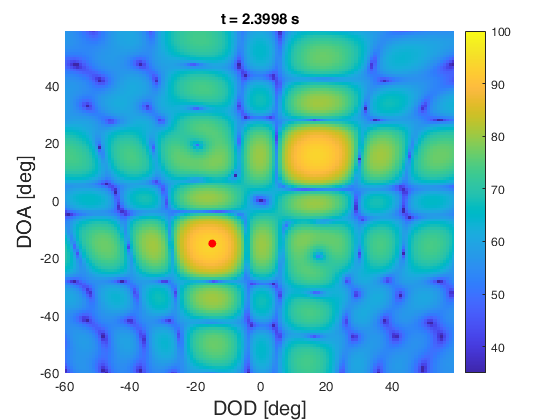}
        \caption{Short baseline with correction.}
        \label{fig:fdm_short_correction}
    \end{subfigure}

    \vspace{0.5em}

    \begin{subfigure}{0.48\textwidth}
        \centering
        \includegraphics[width=\linewidth]{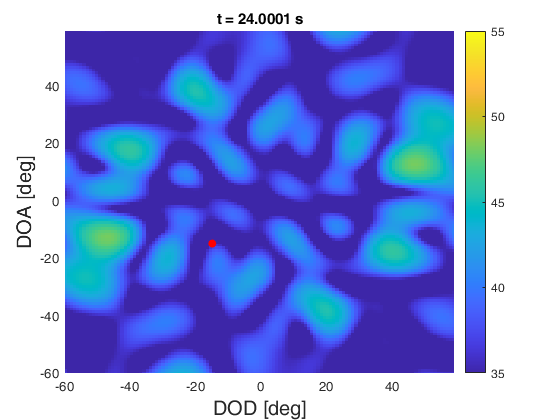}
        \caption{Long baseline without correction.}
        \label{fig:fdm_long_nocorrection}
    \end{subfigure}
    \hfill
    \begin{subfigure}{0.48\textwidth}
        \centering
        \includegraphics[width=\linewidth]{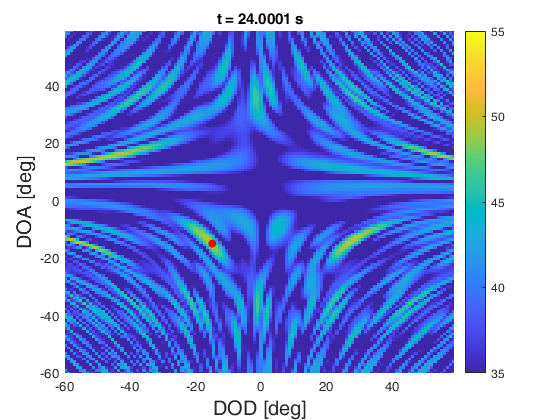}
        \caption{Long baseline with correction.}
        \label{fig:fdm_long_correction}
    \end{subfigure}

    \caption{Comparison of DoD--DoA maps obtained using FDM-based MIMO approach for short- and long-baseline scenarios, with and without correction. The red dot shows the true target location. The results highlight both the uncompensated DoD bias and the distortions introduced by the corrective term for large values of \(B\Delta f\).}
    \label{fig:fdm_comparison}
\end{figure*}

The impact of the FDM correction is illustrated in Fig.~\ref{fig:fdm_comparison}, which compares the DoD--DoA maps for the short- and long-baseline scenarios, with and without correction. The reported frames for different scenarios are selected so that the target lies approximately at the same angular coordinates, allowing a direct comparison of the different effects induced by the baseline length and by the correction procedure. In the long-range scenario, the target signatures are weaker and the output exhibits a lower dynamic range due to the smaller target and the larger propagation distances compared with the short-range case. Therefore, for better visualization, the upper limit of the color scale has been reduced for the long-range maps. 


In the short-baseline case, shown in Fig.~\ref{fig:fdm_comparison}(a)--(b), neglecting the correction produces only a negligible DoD error, and the target remains substantially well localized, also considering the limited angular resolution provided by the small arrays adopted. Applying the correction, while compensating the very small error, leads to minor modifications of the final map. This behavior is expected since the mismatch introduced by FDM scales with the product between the carrier-frequency separation and the baseline length $B\Delta f$. Therefore, for short baselines (and/or small $\Delta f$), the corrective term in \eqref{equtilde} remains limited and does not significantly affect the angular reconstruction.

Conversely, in the long-baseline scenario, reported in Fig.~\ref{fig:fdm_comparison}(c)--(d), the uncompensated phase mismatch becomes substantial. Without correction, the target undergoes a severe DoD mislocation, which in the considered example reaches approximately \(35^\circ\). This confirms that, for large values of \(B\Delta f\), error correction becomes necessary. After the proposed correction, the target peak appears in the expected position. However, the output map reveals important warping effects introduced by the correction itself. In particular, the main lobe of the target response is elongated along an oblique direction, revealing a coupling between the two angular coordinates. Furthermore. the corrected map shows additional ambiguities within the field of view that would not appear in the ideal case since the inter-element spacing at the two arrays is set to be half the wavelength. These distortions are negligible for short baselines, but become clearly visible as the product \(B\Delta f\) increases.

To complete the analysis, we report in Table \ref{tab:resolution_comparison_long_baseline} the DoD and DoA  resolutions obtained for the scenario shown in Fig. \ref{fig:fdm_long_correction}, where the target is located approximately at $\beta_{tx}=\beta_{rx}=-15^\circ$. Specifically, we compare the values obtained with the theoretical expressions in \eqref{eqResFdmDod} and \eqref{eqResFdmDoa} and the values measured from the simulated output. The first row of the table refers to the FDM case and shows a close agreement between the theoretical predictions and the measured resolutions. It is also straightforward to verify that, even with identical transmit and receive arrays, the DoD resolution is compressed relative to the DoA resolution because of the warping effect.  

\subsection{TDM}
\begin{figure*}[!ht]
	\centering
\includegraphics[width =\textwidth]{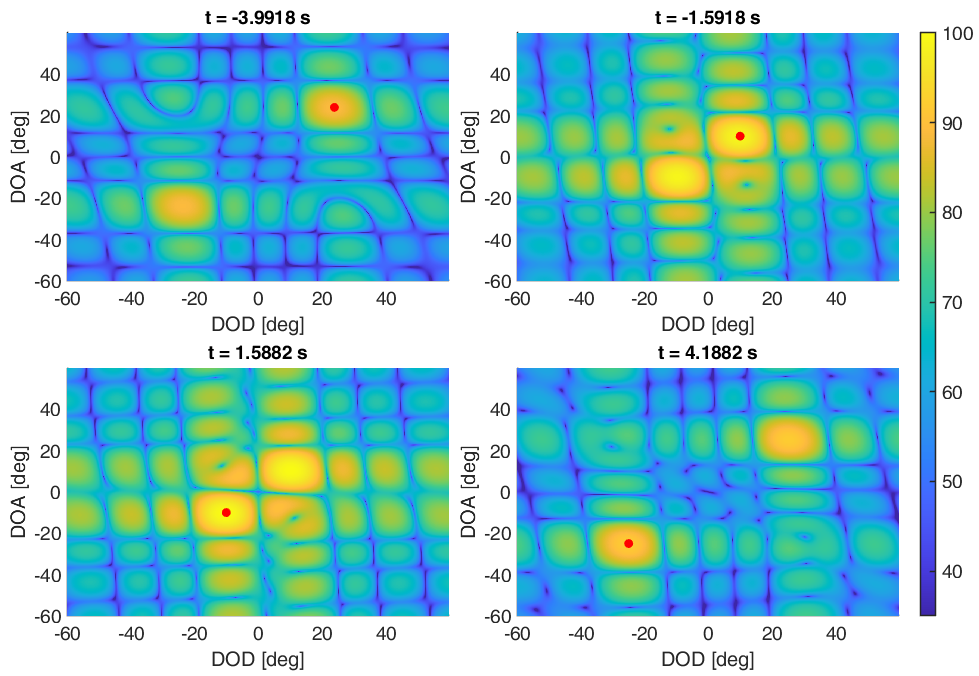}
	\caption{Four frames of DoD-DoA maps over time obtained using MIMO approach for the short baseline scenario using TDM. The red dot shows the true target location}
    \label{fig:tdm_tracking_short}
\end{figure*}

Fig.~\ref{fig:tdm_tracking_short} shows four representative DoD--DoA maps obtained for the short-range scenario at different time instants across the target path when using TDM  and compensating for the motion-induced error. Here we assume the bistatic range rate to be known so that the correction is readily applied. Similarly to the FDM case, the target is correctly localized and tracked over time.

\begin{figure*}[!ht]
    \centering

    \begin{subfigure}{0.49\textwidth}
        \centering
        \includegraphics[width=\linewidth]{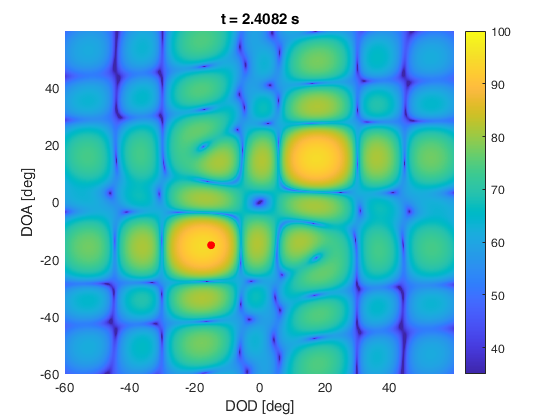}
        \caption{Short baseline without correction}
        \label{fig:tdm_short_nocorrection}
    \end{subfigure}
    \hfill
    \begin{subfigure}{0.49\textwidth}
        \centering
        \includegraphics[width=\linewidth]{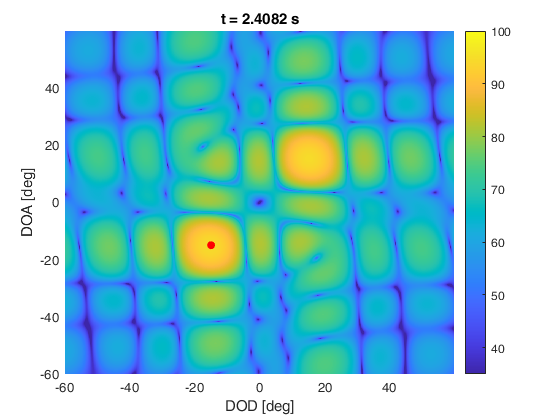}
        \caption{Short baseline with correction}
        \label{fig:tdm_short_correction}
    \end{subfigure}

    \vspace{0.5em}

    \begin{subfigure}{0.49\textwidth}
        \centering
        \includegraphics[width=\linewidth]{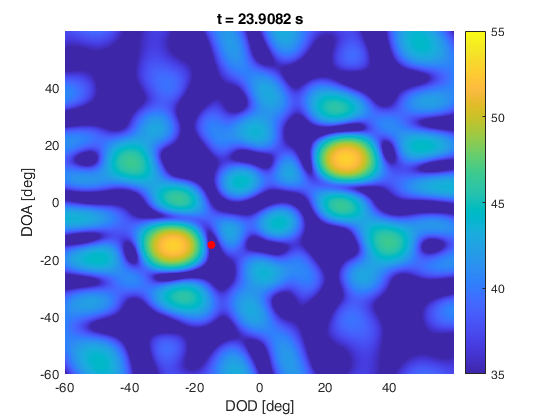}
        \caption{Long baseline without correction}
        \label{fig:tdm_long_nocorrection}
    \end{subfigure}
    \hfill
    \begin{subfigure}{0.49\textwidth}
        \centering
        \includegraphics[width=\linewidth]{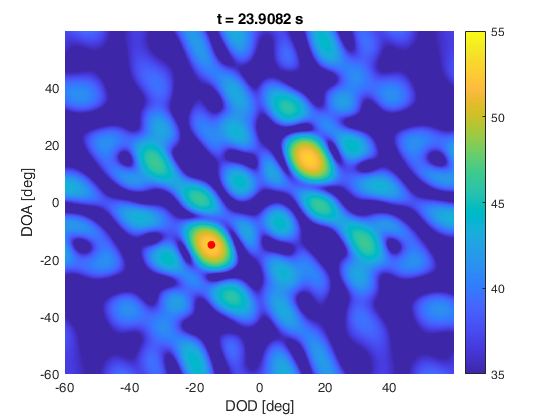}
        \caption{long baseline with correction}
        \label{fig:tdm_long_correction}
    \end{subfigure}

    \caption{Comparison of DoD--DoA maps obtained using TDM-based MIMO approach for short- and long-baseline scenarios, with and without correction. The red dot shows the true target location
    }
    
    \label{fig:tdm_maps_comparison}
\end{figure*}

\begin{figure}[!ht]
        \centering
         \includegraphics[width=\columnwidth]{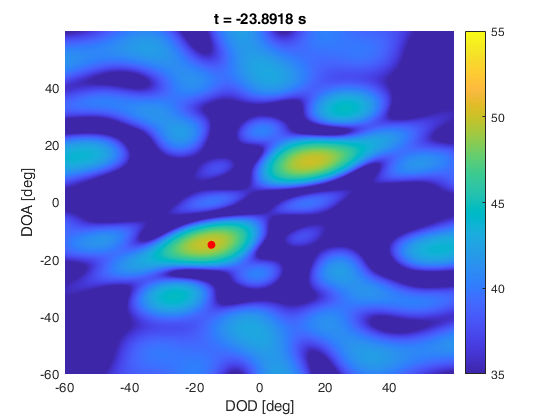}
        \caption{One frame of a DoD-DoA map obtained through MIMO approach using time division technique with correction for short range scenario when the target moves towards positive y-axis.}
        \label{fig:tdm_long_correction_invertedvelocity}
        
\end{figure}

\begin{table}[!ht]
\centering
\caption{Theoretical and measured DoD and DoA resolutions for the long-baseline drone scenario at $t=24$ s, with target location approximately at $\beta_{tx}=\beta_{rx}=-15^\circ$.}
\label{tab:resolution_comparison_long_baseline}
\begin{tabular}{ccccc}
\hline
\textbf{Scheme} & \textbf{DoD theo.} & \textbf{DoD meas.} & \textbf{DoA theo.} & \textbf{DoA meas.} \\ \hline
FDM & 3.63° & 3.28° & 4.89° & 4.69° \\
TDM & 7.80° & 7.50° & 9.93° & 10.31° \\ \hline
\end{tabular}
\end{table}

To better assess the impact of the motion-induced distortive term, Fig.~\ref{fig:tdm_maps_comparison} report DoA/DoD maps for the same cases considered in Fig.~\ref{fig:fdm_comparison}. 

In the short-range scenario (Fig.~\ref{fig:tdm_maps_comparison}(a)--(b)), the difference between corrected and uncorrected maps is almost negligible. Differently from the FDM case, this is not due to the short baseline but to the limited velocity of the human-like target. In fact, given that the product $\dot{R}_\mathrm{bi} \Delta T $ is the relevant parameter for TDM, this result indicates that the distortion induced by the target motion during consecutive transmission slots remains limited under moderate velocities and short inter-slot durations. 
In the long-range case (Fig.~\ref{fig:tdm_maps_comparison}(c)--(d)), the absence of correction produces a more evident DoD estimation error, since for the simulated UAV, the quantity $\dot{R}_{\mathrm{bi}}\Delta T$, which is proportional to the target displacement between consecutive illuminations, is no longer negligible. Nevertheless, with the considered parameters, the resulting angular shift remains significantly smaller than in the corresponding FDM configuration, being approximately \(10^\circ\). For the same reasons, after correction, the target is correctly relocated, and the induced warping remains much more alleviated than that observed in the FDM long-baseline scenario. 

Another observation in Fig. \ref{fig:tdm_maps_comparison}(d), in accordance with the mathematical derivations given in Sec. \ref{sec:TDM}, is that the correction has led to an improved DoD resolution. This is also apparent from the numerical values reported in Table \ref{tab:resolution_comparison_long_baseline} corresponding to Fig. \ref{fig:tdm_long_correction} . 
As an opposite case, an additional result for the same long-range scenario is shown in Fig.~\ref{fig:tdm_long_correction_invertedvelocity}, where the target moves in the opposite direction while the same TDM correction is applied. In this case, reversing the target velocity sign produces a more distorted and widened DoD response. This behavior confirms that, in TDM, the distortion in the map induced by the correction, is influenced not only by the magnitude of the velocity vector but also by its direction.

Overall, the simulation results highlight that the choice between FDM and TDM should be driven by the intended application and operating scenario. In short-range configurations, the distortions introduced by the FDM correction remain negligible, making FDM an attractive solution, particularly for highly dynamic targets. Since all transmitters illuminate the scene simultaneously, FDM does not complicate the TX side e.g. by imposing constraints on the transmission frame duration. In short range scenarios, it enables larger waveform bandwidths to be allocated to each transmitter which is favorable in ISAC-oriented implementations. As the baseline increases, however, the FDM correction becomes progressively more significant, leading to nonlinear warping of the DoD--DoA map and increasingly restrictive array-design constraints. In contrast, the performance of TDM is primarily governed by the product $\dot{R}_{\rm bi}\Delta T$, which remains sufficiently small for a notable range of target velocities, including those associated with typical ground-based targets and UAVs, to limit both the uncompensated angular error and the distortions introduced by the correction. Consequently, TDM represents an attractive solution for medium- and long-range surveillance of such targets. For very high-speed aerial targets observed over long baselines, however, the selection of the preferred multiplexing strategy requires a broader system-level trade-off involving Doppler resolution, update rate, waveform bandwidth, and hardware complexity. In this context, the analytical resolution and ambiguity formulations derived in this work provide useful design guidelines for selecting the most suitable architecture.

\section{Experimental Results}
\label{sec:experimental_results}

This section reports the results of an experimental campaign aimed at demonstrating the feasibility of the proposed MIMO FSR concept in realistic scenarios. The tests performed using the FDM and the TDM strategies are illustrated and discussed in the subsequent sub-sections. For this reason, OFDM signals are also used in lieu of sinusoidal tones in order to verify the suitability of this concept for ISAC applications. 

\subsection{Experimental Results with FDM}
\label{subsec:experimental_results_fdm}
\begin{figure}[!ht]
        \centering
        \includegraphics[width=\columnwidth]{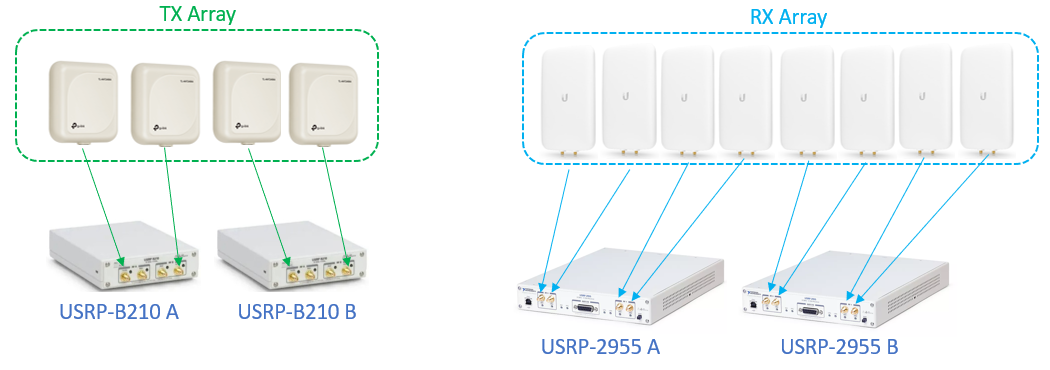}
        \caption{Representation of the acquisition setup for FDM.}
        \label{fig:experimental_setup}
\end{figure}

\begin{figure}[!ht]
        \centering
        \includegraphics[width=\columnwidth]{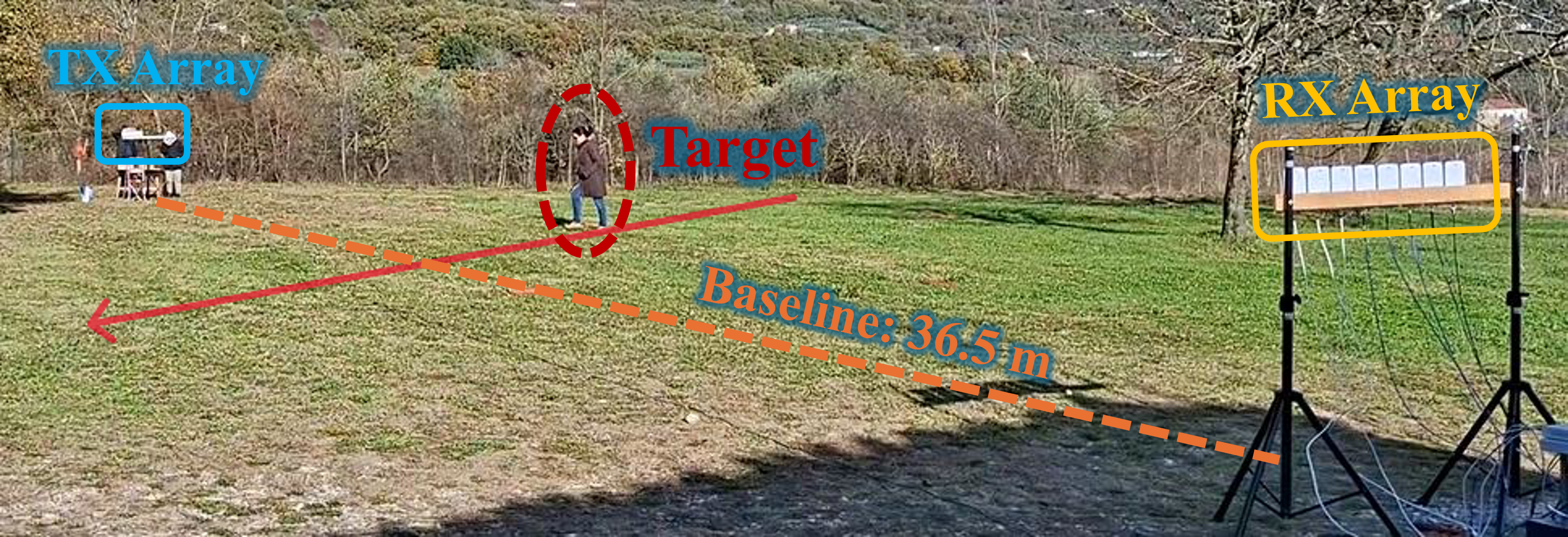}
        \caption{FDM acquisition scenario.}
        \label{fig:foto_configurazione_finale}
\end{figure}

The acquisition setup used to test the FDM-based MIMO-FSR approach is shown in Fig. \ref{fig:experimental_setup}. The performed test is depicted in Fig. \ref{fig:foto_configurazione_finale}  while the main adopted parameters are summarized in Table~\ref{tab:fdm_exp_parameters}. The transmitting array includes $M=4$ TP-LINK TL-ANT2409 antennas, while the receiving array consists of $N=8$ Ubiquiti UMA-D antennas. The transmitted signals are generated using two dual-channel Ettus USRP-B210 devices; they are 16-QAM OFDM waveforms emitted with a carrier frequency spacing $\Delta f=1.5$~MHz around a reference carrier frequency of 2.45~GHz. The four transmitted waveforms occupy adjacent non-overlapped frequency bands, thus allowing the different transmitting contributions to be separated at the receiver side through bandpass filtering. The receiving signals were collected and digitized using two quad-channel NI USRP-2955 devices. Notice that the multi-channel devises used on transmit and on receive are not phase/frequency synchronized since the non-coherent nature of the FSR processing scheme eliminates the need for accurate synchronization. Nevertheless, rough time synchronization and amplitude calibration are achieved by exploiting the direct signals from
the TXs. Finally MATLAB was used for offline processing. 

It is relevant to note that in addition to the spatial DC removal stage, a temporal DC removal is applied immediately beforehand. This helps alleviate contributions from undesired stationary obstacles and mitigate multipath effects, thereby improving the quality of the reconstructed maps.   

\begin{table}[!t]
\centering
\caption{Main parameters of the FDM experimental tests.}
\label{tab:fdm_exp_parameters}
\begin{tabular}{l c}
\hline
\textbf{Parameter} & \textbf{Value} \\
\hline
Carrier frequency & 2.45 GHz \\
Waveform & 16-QAM OFDM \\
FDM frequency separation & $\Delta f = 1.5$ MHz\\
Number of transmitting elements & $M=4$ \\
Transmitting antennas & TP-LINK TL-ANT2409 \\
Transmit antenna gain & 9 dBi \\
Number of receiving elements & $N=8$ \\
Receiving antennas & Ubiquiti UMA-D \\
Baseline length & $B=37$ m \\
TX inter-element spacing & $d_{\mathrm{TX}}=0.98\lambda$ \\
RX inter-element spacing & $d_{\mathrm{RX}}=1.14\lambda$ \\
Target type & Human target \\
Target velocity & Slow walking target \\
\hline
\end{tabular}
\end{table}

\begin{figure*}[!ht]
	\centering
    \includegraphics[width=\textwidth]{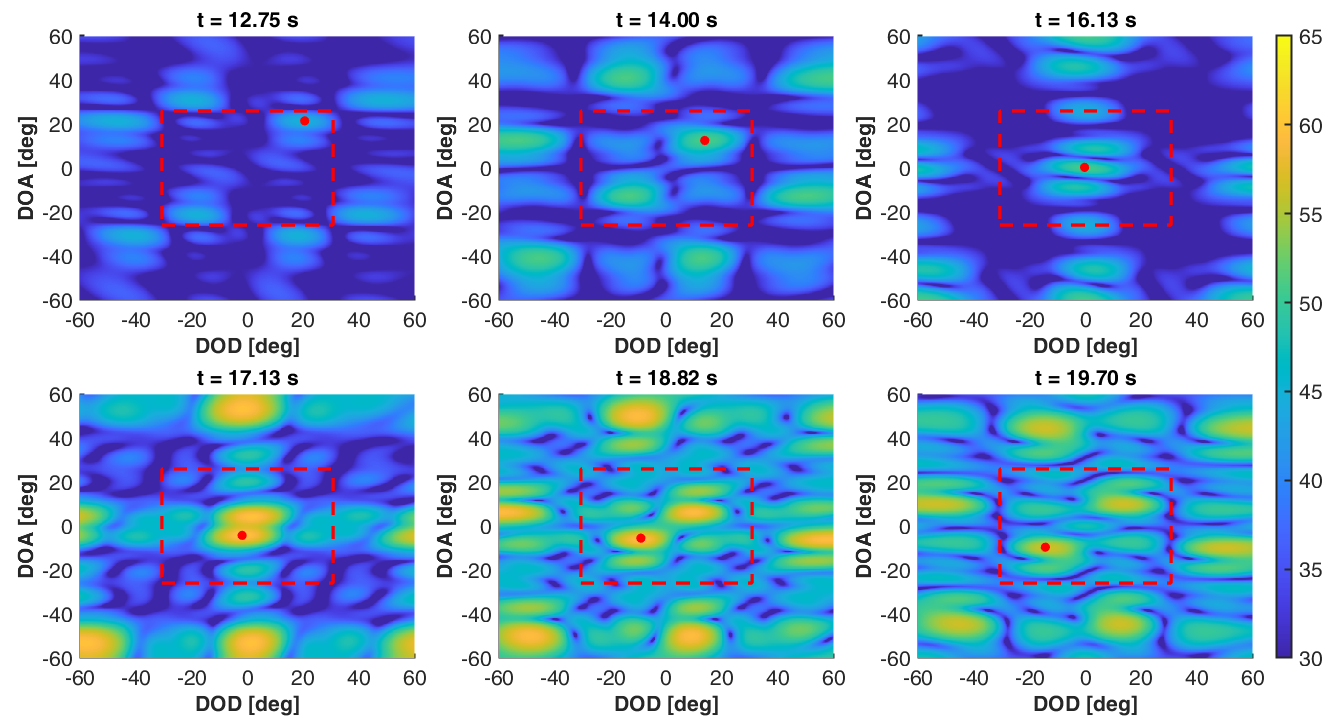}
	\caption{Six frames of DoD-DoA maps over time obtained through MIMO approach for the experimental scenario with FDM. The dashed box denotes the unambiguous area and the red dot shows the true target location.}
    \label{fig:8elements_ofdm}
\end{figure*}

The results are reported in Fig.~\ref{fig:8elements_ofdm}. The figure shows six DoD-DoA maps corresponding to different time instants across the target path; the red circle indicates the ground-truth. These snapshots show that the target instantaneous position can be clearly detected as it approaches, crosses, and then moves away from the baseline.

In addition, each map exhibits several ambiguous peaks. These artifacts arise mainly from the transmit and receive arrays, whose inter-element spacing exceeds $\lambda/2$ due to the physical dimensions of the antenna elements, which precluded a denser array configuration. As a consequence, spatial ambiguities are observed in both the DoD and DoA dimensions.
In all of the maps, the red dashed box denotes the unambiguous angular region, as determined by the same condition \eqref{eqAmbFDM2}.  
Note that the product $\alpha_\mathrm{FDM}R_\mathrm{bi}$ in the considered short-range scenario is very small with respect to $\lambda$ and does not affect the unambiguous region borders. Within this region, as expected, only two dominant lobes are observed, symmetrically located in the first and third quadrants. This phenomenon arises from the sign ambiguity inherent to the adopted non-coherent, amplitude-based FSR approach, whereby 
a target located at an angular position $(\beta_{\mathrm{TX}},\beta_{\mathrm{RX}})$ produces also an indistinguishable response at $(-\beta_{\mathrm{TX}},-\beta_{\mathrm{RX}})$. 
It is worth mentioning that while the corrective term required by the FDM strategy was applied in the processing chain, due to the relatively short baseline and limited frequency separation among the transmitted waveforms in the present experiment, the map distortions remain negligible. 
These results confirm the suitability of the FDM strategy for short-range scenarios and further demonstrate the feasibility of implementing a MIMO-FSR concept within an ISAC framework using OFDM waveforms.

 


\subsection{Experimental Results with TDM}
\label{subsec:experimental_results_tdm}

The experimental tests reported in this section aim to validate the proposed MIMO-FSR concept when the transmitted signals are separated through TDM. The same carrier frequency and baseline length adopted in Section VII.A were considered, while the number of elements in the RX array is reduced to $N=6$. However, the experimental setup was modified to emulate a TDM-MIMO configuration by synthesizing a long transmitting array.

Specifically, due to hardware limitations, real-time generation of multiple transmitted signals in different time slots was not feasible with the available devices. To overcome this limitation, a single transmitting antenna was sequentially moved across a set of equally spaced positions, with signal transmission repeated at each position. This approach enabled the emulation of a synthetic transmitting array composed of $M=14$ elements with half-wavelength inter-element spacing. As a result, no ambiguities are expected in the DoD dimension.
On the other hand, because the synthetic array was generated through successive transmissions, the interval $\Delta T$ between consecutive measurements became extremely large compared to a practical TDM-MIMO system. Consequently, the proof-of-concept experiment had to be conducted under stationary conditions to ensure the validity of the inherent principles.

\begin{figure}[!ht]
        \centering
        \includegraphics[width=\columnwidth]{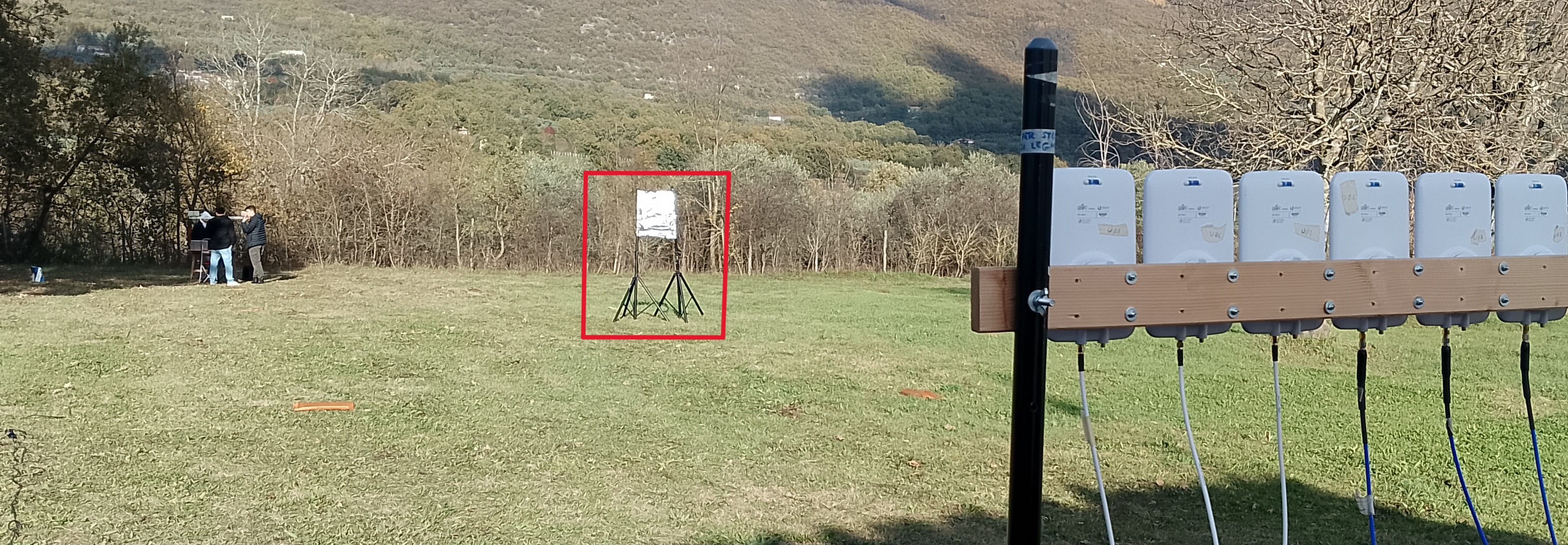}
        \caption{TDM acquisition scenario.}
        \label{fig:tdm_corner_reflector}
\end{figure}

\begin{figure}[!ht]
        \centering
        \includegraphics[width=0.8\columnwidth]{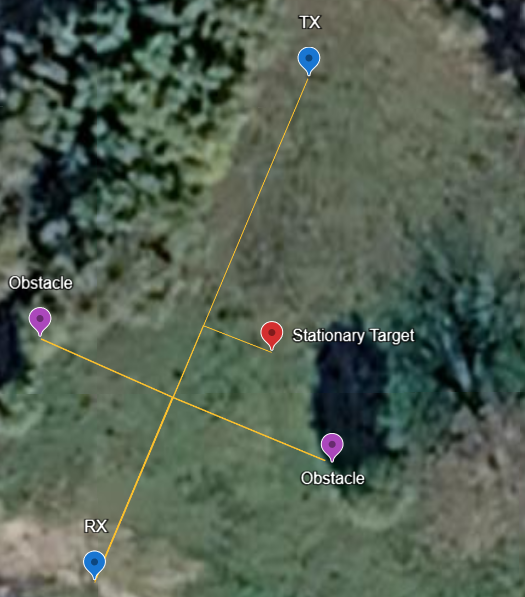}
        \caption{Top view of the acquisition scenario. The TX-RX baseline, the stationary target position and two possible obstacles, corresponding to trees located in the area, are highlighted.}
        \label{fig:tdm_scenario_top_view}
\end{figure}

In detail, the test scenario is shown in Fig. \ref{fig:tdm_corner_reflector} with a top-view reported in Fig. \ref{fig:tdm_scenario_top_view}, while the main parameters of the TDM experimental tests are summarized in Table~\ref{tab:tdm_exp_parameters}. 
The target of interest is a stationary object lying on the central axis of the baseline, at a distance of 5~m from the baseline. The corresponding ground-truth  DoD and DoA coordinates are both equal to $15.32^\circ$. The scene also includes other stationary obstacles, mainly trees, on both sides of the baseline with larger DoA/DoD values. The largest trees have been indicated in Fig. \ref{fig:tdm_scenario_top_view} and are located almost symmetrically with respect to the baseline. Their positions correspond approximately to the angular pair $(\beta_{\mathrm{TX}},\beta_{\mathrm{RX}})=(22.5^\circ,35.6^\circ)$ and to its symmetric counterpart with respect to the baseline. According to the amplitude-based FSR processing, such returns are expected to produce overlapping contributions in the DoD-DoA map.

\begin{table}[!t]
\centering
\caption{Main parameters of the TDM experimental tests.}
\label{tab:tdm_exp_parameters}
\begin{tabular}{l c}
\hline
\textbf{Parameter} & \textbf{Value} \\
\hline
Carrier frequency & 2.45 GHz \\
Waveform & Sinusoidal / 16-QAM OFDM \\
Signal-separation strategy & TDM \\
Number of virtual transmitting elements & $M=14$ \\
TX inter-element spacing & $d_{\mathrm{TX}}=\lambda/2$ \\
Number of receiving elements & $N=6$ \\
RX inter-element spacing & $d_{\mathrm{RX}}=1.14\lambda$ \\
Baseline length & $B=36.5$ m \\
Target type & Stationary Target \\
Target horizontal size & 0.5 m \\
Target vertical size & 0.7 m \\
Target position & $x=5$ m, $y=B/2$ \\
Ground-truth DoD & $15.3215^\circ$ \\
Ground-truth DoA & $15.3215^\circ$ \\
\hline
\end{tabular}
\end{table}

Differently from the FDM case, no corrective term is required in this TDM experiment. Indeed, the corrective term associated with TDM depends on the bistatic velocity of the target, and it vanishes in the stationary case considered here. 
However, the stationary nature of the scenario also affects the DC-removal stage. In particular, a temporal DC-removal (which we apply in the FDM case to alleviate stationary clutter) would suppress the target contribution, since the target response is constant over time. For this reason, the DC-removal was applied only along the spatial dimensions, namely along the transmitting and receiving arrays. In the implemented processing chain, the spatial DC-removal was performed consecutively along the TX and RX dimensions. As a result, the cancellation notch appears at both $\text{DoA}=0^\circ$ and $\text{DoD}=0^\circ$ lines and the stationary target of interest can be preserved along with the other stationary obstacles present in the scene. Furthermore, in contrast to the FDM case we are able to operate also in MISO configuration and perform one-dimensional angular processing in terms of DoD.  

\begin{figure}[!ht]
        \centering
        \includegraphics[width=\columnwidth]{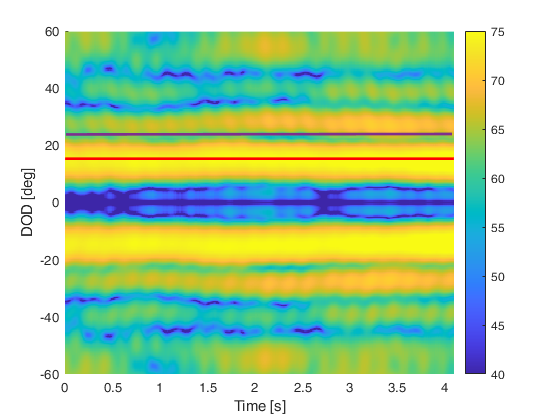}
        \caption{DoD-time map obtained through MISO processing. The solid red line shows the true target DoD while the purple line indicates an approximated location of the stationary obstacle.}
        \label{fig:TDM_DoD_MAP_14tx}
\end{figure}

\begin{figure}[!ht]
        \centering
        \includegraphics[width=\columnwidth]{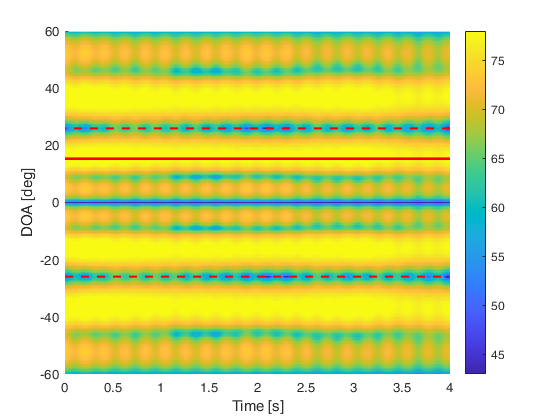}
        \caption{DoA-time map obtained through SIMO processing. The red dashed lines denote the unambiguous angular area and the red solid line shows the true target DoA}
        \label{fig:TDM_DoA_MAP_6rx}
\end{figure}

The first TDM results are reported for the sinusoidal waveform case. Fig.~\ref{fig:TDM_DoD_MAP_14tx} shows the DoD-time map obtained by exploiting the synthetic transmitting array in a MISO-like configuration, namely by jointly processing the signal received from all the transmitting element positions at a single receiving element. Since the target is stationary, its response appears as a persistent vertical signature located at the correct DoD angle, whose ground-truth is indicated by the red line. A weaker contribution, approximately 10~dB below the main target response, can also be observed around the DoD expected for the possible obstacles identified in Fig.~\ref{fig:tdm_scenario_top_view}, approximately indicated by the purple line. Similarly, Fig.~\ref{fig:TDM_DoA_MAP_6rx} reports the DoA-time map obtained through SIMO processing on the receiving array. Also in this case, the target produces a stable signature over time at the expected DoA angle, with the corresponding ground-truth indicated by the red line, while the red dashed lines delimit the unambiguous angular sector. The resulting angular resolution is comparable in the MISO and SIMO cases since in the latter case a smaller number of elements is used but displaced by more than $\lambda/2$. For the same reason, the DoA-time representation also exhibits an additional ambiguous replica of the target return, which appears to mask the contributions of the other stationary obstacles.

\begin{figure}[!ht]
        \centering
        \includegraphics[width=\columnwidth]{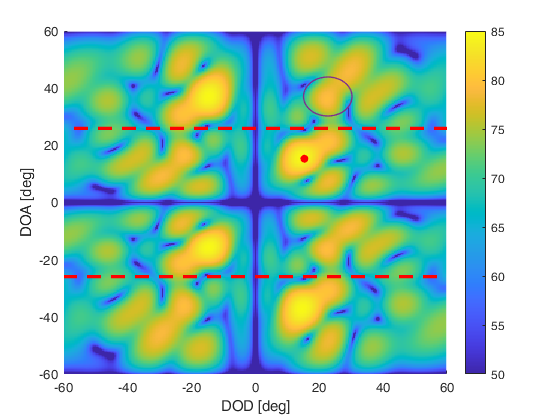}
        \caption{DoD-DoA map obtained using TDM in the stationary-target scenario. The dashed lines denotes the unambiguous area and the red dot shows the true target location. The purple circle indicates the approximate location of the stationary obstacle.}
        \label{fig:TDM_FULL_MAP_14tx_6rx}
\end{figure}

The full MIMO result is shown in Fig.~\ref{fig:TDM_FULL_MAP_14tx_6rx}. Since the scenario is stationary, only one representative frame is reported, as the individual maps are nearly identical, apart from noise and residual fluctuations.

The resulting DoD-DoA map shows a well-defined peak clearly distinguishable from the background. The peak is centered around the expected target position, as confirmed by the red marker representing the ground-truth angular coordinates. This result demonstrates the practical feasibility of the proposed TDM-based MIMO approach. Moreover, consistent with the strategy used to synthesize the TX array, the experiment confirms that the exploited effect remains valid even for widely separated time instants when observing a stationary scene.

In addition to the main peak associated with the target, the DoD–DoA map reveals a weaker response in the angular region corresponding to the potential obstacles highlighted in Fig.~\ref{fig:tdm_scenario_top_view}. This contribution appears about 6--7~dB below the target peak. 
Although it is not possible to unambiguously attribute this response to the trees present in the scene, its angular position is consistent with their estimated DoD-DoA coordinates. This suggests that the proposed processing is also sensitive to weaker stationary scatterers present in the illuminated area at smaller bistatic angles compared to the typical FSR configuration. Incidentally, we observe that these contributions were not visible in the FDM-based MIMO-FSR output since the temporal DC removal only allows the detection of moving targets.

Moreover, as in the FDM case, ambiguous peaks are also present in the TDM DoD-DoA map. The peak in the third quadrant is due to the non-coherent nature of the processing, which makes the response at $(\beta_{\mathrm{TX}},\beta_{\mathrm{RX}})$ indistinguishable from that at $(-\beta_{\mathrm{TX}},-\beta_{\mathrm{RX}})$. The other two replicas are instead associated with the DoA ambiguity introduced by the receiving-array spacing, which is larger than $\lambda/2$.

\begin{figure}[!ht]
        \centering
        \includegraphics[width=\columnwidth]{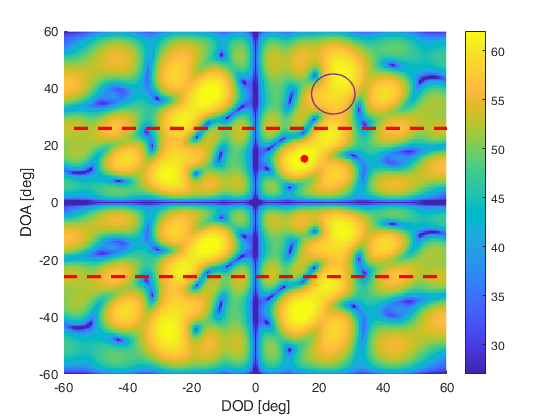}
        \caption{DoD-DoA map obtained using TDM in the stationary-target scenario with OFDM signals. The dashed lines denotes the unambiguous area and the red dot shows the true target location. The purple circle indicates the approximate location of the stationary obstacle.}
        \label{fig:TDM_FULL_MAP_14tx_6rx_ofdm_differents}
\end{figure}

The same TDM approach was also evaluated using 16-QAM OFDM waveforms. A particularly relevant feature of this experiment is that a different OFDM signal realization was transmitted from each virtual TX position; therefore, the OFDM symbols were not kept identical across acquisitions. Despite this additional waveform variability, the resulting DoD–DoA map, shown in Fig.~\ref{fig:TDM_FULL_MAP_14tx_6rx_ofdm_differents}, still exhibits a clear, well-localized peak at the expected target position, along with the returns of the other stationary obstacles.
This result confirms that the effect underlying the proposed non-coherent, amplitude-based MIMO approach depends on the average TX power rather than on the specific waveform transmitted by each antenna. This property is also reflected in the theoretical model, in which distinct indices are assigned to $\tilde{s}_m(t)$. Although the instantaneous OFDM symbols vary across acquisitions, the target response remains spatially coherent after amplitude extraction, low-pass filtering, and spatial DC removal. This finding enhances the flexibility of the proposed approach and its potential to support increased system capacity in ISAC scenarios.


\section{Conclusion}\label{sec:Conclusion}
This paper introduced a generalized MIMO Forward Scatter Radar framework capable of performing joint two-dimensional angular localization through non-coherent, amplitude-only processing. By exploiting antenna arrays at both the transmitter and receiver sides, the proposed approach extends the capabilities of conventional SISO and SIMO FSR systems, enabling simultaneous estimation of the target DoD and DoA without requiring phase synchronization among array elements.

Starting from an ideal signal-separation assumption, a unified spatial processing formulation was derived and subsequently specialized to two practical multiplexing strategies, namely FDM and TDM. For both schemes, the mechanisms responsible for angular distortions were analytically characterized, and dedicated correction procedures were developed. In the FDM case, the correction was shown to depend on the bistatic geometry and to become increasingly important as the product of baseline length and frequency separation grows. In the TDM case, the distortion was shown to depend on target motion through the bistatic range rate, enabling correction either through Doppler side information or integrated three-dimensional processing. Analytical expressions for angular resolution and ambiguity conditions were also derived, revealing the non-uniform effects introduced by the correction procedures.

Numerical simulations demonstrated the effectiveness of the proposed processing schemes in both short-range and long-range scenarios. The results highlighted that uncompensated FDM can lead to severe DoD estimation biases for large baselines, while the associated correction may introduce noticeable map distortions and resolution asymmetries. Conversely, TDM exhibited greater robustness in the considered scenarios, producing smaller angular errors and less pronounced distortions. The theoretical resolution predictions were further validated through simulation, showing good agreement with the measured responses.

The proposed framework was finally verified through real-world experiments using software-defined radio platforms. The FDM campaign demonstrated successful joint DoD–DoA localization of a moving human target in an outdoor forward-scatter configuration, while the TDM campaign confirmed the validity of the proposed processing chain through synthetic-array measurements and stationary-target localization. These experimental results provide practical evidence of the feasibility of MIMO-FSR and confirm its potential for enhanced target localization in forward-scatter environments.

\bibliographystyle{IEEEtran} 
\bibliography{refs} 

@incollection{cherniakov2007basic,
  title={Basic principles of forward-scattering radars},
  author={Cherniakov, M},
  booktitle={Bistatic Radar: Principles and Practice: Part III},
  year={2007},
  publisher={Wiley New York}
}

@article{ustalli2017detection,
  title={Detection performance of a forward scatter radar using a crystal video detector},
  author={Ustalli, Nertjana and Lombardo, Pierfrancesco and Pastina, Debora},
  journal={IEEE Trans. on Aerospace and Electronic Systems},
  volume={54},
  number={3},
  pages={1093--1114},
  year={2017},
  publisher={IEEE}
}

@article{contu2017passive,
  title={Passive multifrequency forward-scatter radar measurements of airborne targets using broadcasting signals},
  author={Contu, Micaela and others},
  journal={IEEE Trans. on Aerospace and Electronic Systems},
  volume={53},
  number={3},
  pages={1067--1087},
  year={2017},
  publisher={IEEE}
}

@article{colone2020dvb,
  title={{DVB-T}-based passive forward scatter radar: Inherent limitations and enabling solutions},
  author={Colone, Fabiola},
  journal={IEEE Trans. on Aerospace and Electronic Systems},
  volume={57},
  number={2},
  pages={1084--1104},
  year={2020},
  publisher={IEEE}
}

@inproceedings{krysik2013gsm,
  title={GSM based passive receiver using forward scatter radar geometry},
  author={Krysik, Piotr and Kulpa, Krzysztof and Samczy{\'n}ski, Piotr},
  booktitle={2013 14th International Radar Symposium (IRS)},
  volume={2},
  pages={637--642},
  year={2013},
  organization={IEEE}
}

@inproceedings{ilioudis2019gnss,
  title={GNSS based passive radar for UAV monitoring},
  author={Ilioudis, Christos V and Cao, Jianlin and Theodorou, Ilias and Striano, Pasquale and Coventry, William and Clemente, Carmine and Soraghan, John},
  booktitle={2019 IEEE Radar Conference (RadarConf)},
  pages={1--6},
  year={2019},
  organization={IEEE}
}

@ARTICLE{Ajorloo2025Multil,
  author={Ajorloo, Abdollah and Qin, Yihua and Colone, Fabiola},
  journal={IEEE Transactions on Aerospace and Electronic Systems}, 
  title={Multichannel Forward Scatter Radar Using Arbitrary Waveforms}, 
  year={2025},
  volume={61},
  number={6},
  pages={17858-17878},
  doi={10.1109/TAES.2025.3610476}}

@inproceedings{myakinkov2003space,
  title={Space-time processing in three-dimensional forward scattering radar},
  author={Myakinkov, AV and Ryndyk, AG},
  booktitle={4th International Conference on Antenna Theory and Techniques (Cat. No. 03EX699)},
  volume={1},
  pages={355--358},
  year={2003},
  organization={IEEE}
}

@inproceedings{Qin2024,
	title={Feasibility Study for a Multichannel Forward Scatter Radar Exploiting Amplitude-based Array Processing},
	author={Qin, Y and Ajorloo, A and Colone, F},
	booktitle={International Conference Radar 2024 (forthcomming)},
	year={2024},
	organization={IEEE}
}

@inproceedings{ajorloo2024preliminary,
  title={Preliminary Experimental Results for a Multi-channel Forward Scatter Radar},
  author={Ajorloo, Abdollah and Qin, Yihua and Bongioanni, Carlo and Colone, Fabiola},
  booktitle={2024 IEEE International Workshop on Technologies for Defense and Security (TechDefense)},
  pages={496--501},
  year={2024},
  organization={IEEE}
}

@inbook{Gashinova2017,
	title = {Forward scatter radar},
	author = {Gashinova, M and Daniel, L and Myakinkov, A and Cherniakov M.},
	booktitle = {Novel Radar Techniques and Applications Volume 1: Real Aperture Array Radar, Imaging Radar, and Passive and Multistatic Rada},
	year={2017},
	publisher = {Institution of Engineering and Technology},
}

@inproceedings{ustalli2019target,
	title={Target kinematic parameters estimation in single input multi output Forward Scatter Radar configuration},
	author={Ustalli, Nertjana and Pastina, Debora and Lombardo, Pierfrancesco},
	booktitle={2019 IEEE Radar Conference (RadarConf)},
	pages={1--6},
	year={2019},
	organization={IEEE}
}

@article{pastina2016target,
	title={Target motion estimation via multi-node forward scatter radar system},
	author={Pastina, Debora and Contu, Micaela and Lombardo, Pierfrancesco and Gashinova, Marina and De Luca, Alessandro and Daniel, Liam and Cherniakov, Mikhail},
	journal={IET Radar, Sonar \& Navigation},
	volume={10},
	number={1},
	pages={3--14},
	year={2016},
	publisher={Wiley Online Library}
}

@inproceedings{rabaste2013signal,
	title={Signal waveforms and range/angle coupling in coherent colocated MIMO radar},
	author={Rabaste, Olivier and Savy, Laurent and Cattenoz, Mathieu and Guyvarch, Jean-Paul},
	booktitle={2013 International Conference on Radar},
	pages={157--162},
	year={2013},
	organization={IEEE}
}

@INPROCEEDINGS{11046028,
	author={Qin, Yihua and Ajorloo, Abdollah and Colone, Fabiola},
	booktitle={2025 26th International Radar Symposium (IRS)}, 
	title={MUSIC Algorithm for Amplitude-Based Array Processing in Multi-Channel Forward Scatter Radar}, 
	year={2025},
	volume={},
	number={},
	pages={1-8},
	doi={10.23919/IRS64527.2025.11046028}}

@inproceedings{ajorloo2025forward,
  title={Forward Scatter Radar Using a MIMO Configuration},
  author={Ajorloo, Abdollah and Colone, Fabiola},
  booktitle={2025 IEEE Radar Conference (RadarConf25)},
  pages={757--762},
  year={2025},
  organization={IEEE}
}

@article{qin2025music,
  title={MUSIC-Based Multi-Channel Forward-Scatter Radar Using OFDM Signals},
  author={Qin, Yihua and Ajorloo, Abdollah and Colone, Fabiola},
  journal={Sensors},
  volume={25},
  number={24},
  pages={7621},
  year={2025},
  publisher={MDPI}
}

@inproceedings{Ajorloo2026Eurad,
	title={Preliminary Experimental Results of a {MIMO} Forward Scatter Radar using Frequency Division},
	author={Ajorloo, A and Ciciretti, S. A and Zorzi, M and Bongioanni, C and Colone, F},
	booktitle={2026 23nd European Radar Conference ((forthcomming))},
	year={2026},
	organization={IEEE}
}

@article{gashinova2013phenomenology,
	title={Phenomenology of Doppler forward scatter radar for surface targets observation},
	author={Gashinova, Marina and Daniel, Liam and Sizov, Vladimir and Hoare, Edward and Cherniakov, Mikhail},
	journal={IET Radar, Sonar \& Navigation},
	volume={7},
	number={4},
	pages={422--432},
	year={2013},
	publisher={Wiley Online Library}
}

@book{Willis2005,
	title={Bistatic radar},
	author={Willis, Nicholas J},
	volume={2},
	year={2005},
	publisher={SciTech Publishing}
}

@article{Glaser1985,
	title={Bistatic RCS of complex objects near forward scatter},
	author={Glaser, Jerome I},
	journal={IEEE Transactions on Aerospace and Electronic Systems},
	number={1},
	pages={70--78},
	year={1985},
	publisher={IEEE}
}

\end{document}